\documentclass{ws-ijmpa}
\usepackage[super,compress]{cite}
\usepackage{graphicx}
\begin{document}

\markboth{Authors' Names}
{Instructions for Typing Manuscripts (Paper's Title)}

%
\catchline{}{}{}{}{}
%

\title{IS GAUGE-INVARIANT COMPLETE DECOMPOSITION OF THE NUCLEON SPIN
POSSIBLE ?}

\author{MASASHI WAKAMATSU}


\address{Department of Physics, Faculty of Science, Osaka University\\
Toyonaka, Osaka 560-0043,
Japan\\
wakamatu@phys.sci.osaka-u.ac.jp}



\maketitle

\begin{history}
\received{Day Month Year}
\revised{Day Month Year}
\end{history}

\begin{abstract}
Is gauge-invariant complete decomposition of the nucleon spin possible?
Although it is a difficult theoretical question which has not
reached a complete consensus yet, a general agreement now is that
there are at least two physically inequivalent gauge-invariant
decompositions (I) and (II) of the nucleon. 
In these two decompositions, the intrinsic spin parts of quarks
and gluons are just common. What discriminate these two decompositions 
are the orbital angular momentum parts.
The orbital angular momenta of quarks and gluons appearing in the
decomposition (I) are the so-called ``mechanical'' orbital angular
momenta, while those appearing in the decomposition (II) are the 
generalized (gauge-invariant) ``canonical'' ones.
By this reason, these decompositions are also called the
``mechanical'' and ``canonical'' decompositions of the nucleon spin,
respectively. A crucially important question is which decomposition
is more favorable from the observational viewpoint. The main objective
of this concise review is to try to answer this question with careful
consideration of recent intensive researches on this problem.  
\keywords{nucleon spin decomposition; gauge invariance; observability.}
\end{abstract}

\ccode{PACS numbers:12.38.-t, 12.20.-m, 14.20.Dh, 03.50.De}

\tableofcontents

\section{Introduction}
The so-called ``nucleon spin puzzle'' raised by the epoch-making
measurements by the EMC Collaboration in 1987 is still one of the most
fundamental problems in quantum chromodynamics
(QCD) \cite{EMC88,EMC89}. (General review of the nucleon spin
problem, can, for example, be found in 
Refs.~\refcite{AEL95,HY_Chen96,LR00,FJ01,Bass05,KCL09,BMN10,ABHM13}.)
In the past ten years, there have been several remarkable progresses
on this problem from the experimental side.
First, a lot of experimental evidence has been accumulated, which
indicates that the gluon polarization inside the nucleon would not be
extremely large \cite{COMPASSG06,PHENIX06,STAR06A,STAR06B}.
At the least, now it seems widely
accepted that the $U_A (1)$-anomaly motivated explanation
of the nucleon spin puzzle \cite{AR88,CCM88,ET88}
is disfavored. Second, the quark
spin fraction or the net longitudinal quark polarization $\Delta q$
has been fairly precisely determined through high-statistics
measurements of the deuteron spin structure function by
COMPASS \cite{COMPASS05,COMPASS07} and the HEREMES group \cite{HERMES07}.
According to these analyses, the portion of the nucleon spin coming
from the intrinsic quark spin turned out to be around 1/3. 
What carries the rest 2/3 of the nucleon spin, then ? This is a fundamental
question of QCD, which we must answer.
To answer this question unambiguously, we cannot avoid to clarify the
following issues :  

\begin{itemlist}
\item What is a precise (QCD) definition of each term of the decomposition ?
\item How can we extract individual term by means of  direct measurements ?
\end{itemlist}

Since QCD is a color SU(3) gauge theory, and because the general principle
of physics dictates that gauge-invariance is a necessary condition of
observability, the gauge-invariance is believed to play a crucially important
role in the nucleon spin decomposition
problem. \cite{JM90,Ji97PRL,Ji98,Teryaev98,BJ99,SW00,BLT04}
In the past several years, this interesting but difficult problem has
been an object of intense debate 
\cite{Chen08,Chen09,Wong10,Wang10,Sun10,Chen11A,Chen11B,Goldman11,Chen12,
Tiwari08,Tiwari13,Ji11A,Ji11B,WakaSD10,WakaSD11A,WakaSD11B,WakaSD11C,WakaSD12,
WakaSD13,Leader11,Leader12,LL12,Leader13,Lorce12,Lorce13A,Lorce13B,Lorce13C,
Lorce14,Hatta11,Hatta12,HY12,HTY13,BBC09,Burkardt05,Burkardt13,Cho11A,Cho11B,Cho12,
ZhangPak12,Pak_Zhang12,GS12,GS13,ZH11,ZHH13,LinLiu12,JXZ12,JXY12A,JXY12B,
JZZ13,Ji13,HJZ13,XJZZ13,HKMR13,HKM14}.
Very recently, Leader and Lorc\'{e} wrote a fairly extensive first review,
which nicely summarizes the current status of the nucleon spin
decomposition problem \cite{Leader_Lorce13}.
The purpose of the present shorter
review is to give a general survey on the same problem
from somewhat different viewpoint from theirs.

The broad guideline of the paper is as follows. First, in Sect.2, we start with
concisely reviewing the history of the nucleon spin decomposition problem.
It is explained why there exist two totally different decompositions of the
nucleon spin and how they are different.  
Next, in Sect.2, we shall dwell on controversial theoretical issues on
the gauge-invariant decomposition problem of the nucleon spin.
We point out that there are conceptually opposing two approaches to the problem.
The one is the standard gauge-fixing approach, while the other is the so-called
gauge-invariant-extension (GIE) approach. A critical difference between
these two ways of thinking is explained in detail.
In Sect. 4, we briefly introduce the recent controversies on
the transverse nucleon spin decomposition. The main focus there
is put on clarifying a significant difference between the transverse spin
sum rule and the longitudinal spin sum rule.
In Sect.5, we shall look at the question whether the canonical orbital
angular momentum (OAM) truly satisfies the standard SU(2) commutation relation.
Contrary to wide-spread
belief, it is shown that neither the canonical OAM nor the intrinsic
spin of the massless gauge field satisfies the SU(2) commutation
relation, and that this fact is inseparably connected with
the vanishing mass of the photon or the gluon. An important conclusion
drawn from this consideration is that only the longitudinal component
of the total gluon angular momentum
can be decomposed into its intrinsic spin and orbital OAM parts in
a gauge- and frame-independent way. What characterizes the difference
between the two different OAMs of the quarks and gluons is the potential
angular momentum in the terminology of Ref.~\refcite{WakaSD10}.
To understand its physical meaning is of vital importance
for answering the question ``Which of the canonical OAM or the mechanical
OAM is a {\it physical} quantity from the viewpoint of observability?"
Naturally, only the high-energy deep-inelastic-scattering (DIS) measurements,
which raised the
problem in the first place, would provide us with a possible practical means
to answer the proposed question on the nucleon spin decompositions.
Section 7 is therefore devoted to the most important issue of the nucleon
spin decomposition problem : 
how can we relate the two existing nucleon spin decompositions, i.e.
the mechanical decomposition and the canonical decomposition, to
direct high-energy deep-inelastic-scattering (DIS) observables?
We try to answer this question with the help of several recent researches
in this direction.
Finally, in Sect.8, we shall summarize what we have leaned and make
some concluding remarks.

\section{A Brief History of the Nucleon Spin Decomposition Problem}

\begin{figure}[h]
\begin{center}
\includegraphics[width=8.0cm]{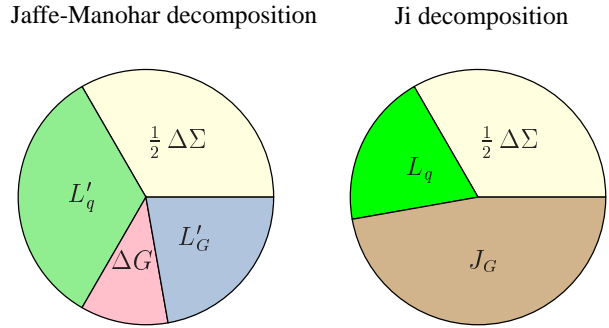}
\caption{\label{Fig1}Two widely-known nucleon spin decompositions.}
\end{center}
\end{figure}

The existence of two different decompositions of the nucleon spin has
been long known in the QCD spin physics community. One is the Jaffe-Manohar
decomposition given as \cite{JM90}
\begin{eqnarray}
 \mbox{\boldmath $J$}_{QCD} &=&
 \int \psi^\dagger \,\frac{1}{2} \, 
 \mbox{\boldmath $\Sigma$} \,
 \psi \,d^3 x \ + \ 
 \int \psi^\dagger \,\mbox{\boldmath $x$} \times
 \frac{1}{i} \,
 \nabla \,\psi \,d^3 x \nonumber \\ 
 &+& \int \mbox{\boldmath $E$}^a \times
 \mbox{\boldmath $A$}^a \,d^3 x \ + \ 
 \int E^{a i} \,\mbox{\boldmath $x$} \times 
 \nabla \,A^{a i} \,d^3 x ,
\end{eqnarray}
with $a$ being the color index of the gluon field, while the other is
the Ji decomposition given as follows \cite{Ji97PRL,Ji98} :
\begin{eqnarray}
 \mbox{\boldmath $J$}_{QCD} &=&
 \int \psi^\dagger \,\frac{1}{2} \,
 \mbox{\boldmath $\Sigma$} \,
 \psi \,d^3 x \ + \ 
 \int \psi^\dagger \,\mbox{\boldmath $x$} \times
 \frac{1}{i} \,
 \mbox{\boldmath $D$} \,
 \psi \,d^3 x \nonumber \\
 &+& \int \mbox{\boldmath $x$} \times
 (\mbox{\boldmath $E$}^a \times
 \mbox{\boldmath $B$}^a ) \,d^3 x , \ \ \ 
\end{eqnarray}
where $\bm{D}$ is the standard covariant derivative defined by 
$\bm{D} \equiv \nabla - i \,g \,\bm{A}$.
In these popular decompositions, only the
intrinsic quark spin part is common, and the other parts are
all different. (See Fig.\ref{Fig1}.)
An apparent disadvantage of the Jaffe-Manohar decomposition is that
each term is not separately gauge-invariant, except for the quark spin part. 
On the other hand, each term of the Ji decomposition is
separately gauge-invariant. Unfortunately, it was claimed
and has been long believed that further gauge-invariant
decomposition of the total gluon angular momentum into its spin and
orbital parts is impossible in this widely-known gauge-invariant decomposition.

An especially annoying observation was that, since the quark orbital
angular momenta (OAMs) in the two decompositions are apparently different, i.e.
\begin{equation}
 L^\prime_q \ \neq \ L_q ,
\end{equation}
one must inevitably conclude that the sum of the gluon spin and the gluon
OAM in the Jaffe-Manohar decomposition
does not coincide with the total gluon angular momentum in the Ji decomposition,
\begin{equation}
 \Delta G \ + \ L^\prime_G \ \neq \ J_G .
\end{equation}

Intensive debates began several years ago when
Chen et al. \cite{Chen08,Chen09} proposed
a new gauge-invariant decomposition of nucleon spin.
The basic idea is to decompose the gluon field $\mbox{\boldmath $A$}$ into
two parts, i.e. the physical component $\mbox{\boldmath $A$}_{phys}$ and
the pure-gauge one $\mbox{\boldmath $A$}_{pure}$.
Under general gauge transformations $U(x)$, the physical part is
supposed to transform covariantly,
\begin{equation}
 \bm{A}_{phys} (x) \ \rightarrow \ \bm{A}^\prime (x) \ = \ 
 U(x) \,\bm{A}_{phys} (x) \,U^\dagger (x),
\end{equation}
while the pure-gauge part is required to transform inhomogeneously, i.e. as
\begin{equation}
 \bm{A}_{pure} (x) \ \rightarrow \ \bm{A}^\prime_{pure} (x) \ = \ 
 U(x) \,\left(\, \bm{A}_{pure} (x) \ + \ \frac{i}{g} \,\nabla \,\right) \,
 U^\dagger (x).
\end{equation}
We recall that, in the case of quantum electrodynamics (QED), their decomposition
is nothing but the familiar transverse-longitudinal decomposition of
the vector potential $\bm{A} (x)$ given as
\begin{equation}
 \bm{A} (x) \ = \ \bm{A}_\perp (x) \ + \ \bm{A}_\parallel (x),
\end{equation}
with the conditions :
\begin{equation}
 \nabla \cdot \bm{A}_\perp (x) \ = \ 0, \hspace{10mm} 
 \nabla \times \bm{A}_\parallel (x) \ = \ 0 .
\end{equation}
As is well-known, this decomposition is {\it unique} owing to the famous
Helmholz theorem, once the Lorentz frame of reference is
fixed.\footnote{To be more precise, the uniqueness of the decomposition
is guaranteed by a supplemental condition that the gauge field $\bm{A}$
falls off faster than $1 / r^2$ at the spatial infinity.
This condition is not necessarily satisfied in some
singular gauges like the light-cone gauge.} 
In the case of quantum chromodynamics (QCD), a similar decomposition
is not so simple as the abelian case. 
To uniquely specify the decomposition, Chen et al. impose an additional
condition for the physical component,
\begin{equation}
 [\bm{A}_{phys}, \bm{E}] \ \equiv \ \bm{A}_{phys} \cdot \bm{E} \ - \ 
 \bm{E} \cdot \bm{A}_{phys} \ = \ 0 ,
\end{equation}
where $\bm{E}$ is the color electric field.
(An alternative and simpler condition for fixing the physical component was
proposed later by themselves \cite{Chen11A} and also by
Zhou, Huang and Huang \cite{ZHH13} as a non-Abelian generalization
of the transversality condition $\nabla \cdot \bm{A}_{phys} = 0$ for
the Abelian vector potential. It is given by
$\bm{D}_{pure} \cdot \bm{A} = 0$ with $\bm{D}_{pure} \equiv 
\nabla - i \,g \,[\bm{A}_{pure}, \,\cdot \,]$ being the pure-gauge
covariant derivative for the adjoint representation.) 
In any case, after imposing such an additional
condition, which is supposed to uniquely fix the decomposition of the gluon
field $\bm{A}$ into $\bm{A}_{phys}$ and $\bm{A}_{pure}$, Chen et al.
proposed the following decomposition of the nucleon spin : 
\begin{eqnarray}
 \mbox{\boldmath $J$}_{QCD} \ = \  
 \mbox{\boldmath $S$}^\prime_{q} \ + \ 
 \mbox{\boldmath $L$}^\prime_{q} \ + \ 
 \mbox{\boldmath $S$}^\prime_{G} \ + \ 
 \mbox{\boldmath $L$}^\prime_{G} ,
\end{eqnarray}
where
\begin{eqnarray}
 \mbox{\boldmath $S$}^\prime_q &=& \int \,\psi^\dagger \,
 \frac{1}{2} \,\mbox{\boldmath $\Sigma$} \,\psi \,d^3 x , \\
 \mbox{\boldmath $L$}^\prime_q &=& \int \,\psi^\dagger \,
 \mbox{\boldmath $x$} \times
 \left(\,\frac{1}{i} \,\nabla \ - \ g \,\mbox{\boldmath $A$}_{pure} \,\right)
 \,\psi \,d^3 x , \\
 \mbox{\boldmath $S$}^\prime_G &=& \int \,
 \mbox{\boldmath $E$}^a \times 
 \mbox{\boldmath $A$}^a_{phys} \,d^3 x , \\
 \mbox{\boldmath $L$}^\prime_G &=& \int \,E^{aj} \,
 (\mbox{\boldmath $x$} \times \bm{D}_{pure}) \,
 A^{aj}_{phys} \,d^3 x ,
\end{eqnarray}
with $\bm{D}_{pure} = \nabla - i \,g \,\bm{A}_{pure}$.
A remarkable feature of this decomposition is that each term is
separately gauge-invariant, as can easily be verified from the
above-mentioned covariant and inhomogeneous gauge transformation
properties of the physical and pure-gauge components of the gluon.
Also noteworthy is that it reduces to
the {\it gauge-variant} decomposition of Jaffe and Manohar in a particular
gauge, $\mbox{\boldmath $A$}_{pure} = 0$, and $\mbox{\boldmath $A$} = 
\mbox{\boldmath $A$}_{phys}$.

Soon after, however, Wakamatsu pointed out that the way of gauge-invariant
complete decomposition of nucleon spin is not necessarily unique,
and proposed yet another decomposition of the
nucleon spin given by \cite{WakaSD10}
\begin{eqnarray}
 \mbox{\boldmath $J$}_{QCD} \ = \  
 \mbox{\boldmath $S$}_{q} \ + \ 
 \mbox{\boldmath $L$}_{q} \ + \ 
 \mbox{\boldmath $S$}_{G} \ + \ 
 \mbox{\boldmath $L$}_{G} ,
\end{eqnarray}
where
\begin{eqnarray}
 \mbox{\boldmath $S$}_q &=& \int \,\psi^\dagger \,
 \frac{1}{2} \,\mbox{\boldmath $\Sigma$} \,\psi \,d^3 x , \\
 \mbox{\boldmath $L$}_q &=& \int \,\psi^\dagger \,
 \mbox{\boldmath $x$} \times
 \left(\,\frac{1}{i} \,\nabla \ - \ g \,\mbox{\boldmath $A$} \,\right)
 \,\psi \,d^3 x , \\
 \mbox{\boldmath $S$}_G &=& \int \,
 \mbox{\boldmath $E$}^a \times 
 \mbox{\boldmath $A$}^a_{phys} \,d^3 x , \\
 \mbox{\boldmath $L$}_G &=& \int \,E^{aj} \,
 (\mbox{\boldmath $x$} \times \bm{D}_{pure}) \,
 A^{aj}_{phys} \,d^3 x \ + \ 
 \int \,\rho^a \,(\mbox{\boldmath $x$} \times 
 \mbox{\boldmath $A$}^a_{phys} ) \,d^3 x . \label{Eq-Lg}
\end{eqnarray}
The characteristic features of this decomposition are as follows.
First, the quark parts of this decomposition (both of spin and
orbital parts) are just common with the Ji decomposition.
Second, the quark and gluon intrinsic spin parts are common
with the Chen decomposition. A crucial difference with the Chen
decomposition resides in the orbital parts.
Namely, although the sums of the quark and gluon OAMs in the two
decompositions are the same, i.e.
\begin{equation}
 \mbox{\boldmath $L$}_q \ + \ \mbox{\boldmath $L$}_G
 \ = \  
 \mbox{\boldmath $L$}^\prime_q \ + \ 
 \mbox{\boldmath $L$}^\prime_G ,
\end{equation}
each term is different in such a way that 
\begin{equation}
 \mbox{\boldmath $L$}_G - \mbox{\boldmath $L$}^\prime_G
 \ = \ 
  - \,
 (\,\mbox{\boldmath $L$}_q - \mbox{\boldmath $L$}^\prime_q)
 \ = \ 
 \int \,\rho^a \,(\mbox{\boldmath $x$} \times 
 \mbox{\boldmath $A$}^a_{phys}) \,d^3 x \ \equiv \ 
 \bm{L}_{pot}.
\end{equation}
The difference arises from the treatment of the 2nd term of Eq.(\ref{Eq-Lg}).
He call this term the {\it potential angular momentum} $\bm{L}_{pot}$,
because the QED correspondent of this term is the orbital angular momentum
(OAM) carried by the electromagnetic field or potential.\footnote{This is just the
quantity appearing in the Feynman paradox raised in his famous textbook of
classical electrodynamics \cite{BookFeynman65}.}
Wakamatsu includes this term in the {\it gluon} OAM part,
while Chen et al. include it in the {\it quark} OAM part.

To understand the difference more clearly, let us first recall the
fact that the potential angular momentum term can also be expressed as
\begin{equation}
 \int \,\rho^a (\mbox{\boldmath $x$} \times \mbox{\boldmath $A$}^a_{phys})
 \,d^3 x \ = \ g \,\int \,\psi^\dagger \,
 \mbox{\boldmath $x$} \times \mbox{\boldmath $A$}_{phys} \,\psi \,d^3 x .
\end{equation}
Note that this term is solely gauge-invariant, as can easily be convinced from
the covariant (or homogeneous) gauge transformation property of the physical
part of the gluon field $\mbox{\boldmath $A$}_{phys}$.
This means that the gauge principle alone cannot say in which part of the
decomposition, one should include the potential angular momentum term.
One certainly has a freedom to include it into the quark OAM part as well,
which would lead to the Chen decomposition.
In fact, if one adds the potential angular momentum to
the quark OAM term of the Ji (or Wakamatsu) decomposition,
the physical part of $\bm{A}$ is exactly canceled out and the pure gauge
part is left, which just leads to the quark OAM term of the Chen
decomposition in the following manner :  
\begin{eqnarray}
 &\,& \mbox{\boldmath $L$}_q \,(\mbox{Ji or Wakamatsu}) 
 \ \ + \ \ \mbox{\boldmath $L$}_{pot}
 \nonumber \\
 &=&
 \int \,\psi^\dagger \,\mbox{\boldmath $x$} \times 
 \left(\,\frac{1}{i} \,\nabla \,- \,g \,\mbox{\boldmath $A$} \,\right) \,
 \psi \,d^3 x \ + \ g \,\int \,\psi^\dagger \,
 \mbox{\boldmath $x$} \times \mbox{\boldmath $A$}_{phys} \,
 \psi \,d^3 x \nonumber \\
 &=&
 \int \,\psi^\dagger \,\mbox{\boldmath $x$} \times 
 \left(\,\frac{1}{i} \,\nabla \,- \,g \,\mbox{\boldmath $A$}_{pure} \,\right) \,
 \psi \,d^3 x \ \ \  = \ \ \ 
 \mbox{\boldmath $L$}^{\prime}_q \,(\mbox{Chen}) .
\end{eqnarray}

It seems true that the two complete decompositions of the nucleon spin, i.e. the
one due to Chen et al. and the other due to Wakamatsu, are both gauge-invariant.
However, a disadvantage of these decompositions is that they are given in
noncovariant forms.
This is not convenient, for example, if one tries to connect these
decompositions with high-energy DIS observables.
Also from more general viewpoint, the non-covariant treatment makes it hard
to check out the Lorentz-frame dependence or independence of the nucleon spin
sum rule derived on the basis of them.
The ``seemingly" covariant generalization of the gauge-invariant decomposition
was given by Wakamatsu \cite{WakaSD11A}.\footnote{The word ``seemingly" here
means that the decomposition $A^\mu = A^\mu_{phys} + A^\mu_{pure}$, although being
covariantly-looking, is intrisically noncovariant,
since it is essentially a transverse-longitudinal decomposition, while what is transverse depends on the choice of Lorentz-frame of reference.}
The starting point of this proposal is
the formally covariant decomposition of the full gauge field $A^\mu$ into
its physical and pure-gauge parts, i.e. 
$A^\mu = A^\mu_{phys} + A^\mu_{pure}$. 
He showed that the gauge-invariant decomposition of the QCD angular momentum
tensor can be obtained with use of the following three conditions only.
The first is the pure-gauge condition for $A^\mu_{pure}$, 
\begin{eqnarray}
 F^{\mu \nu}_{pure} \ \equiv \ 
 \partial^\mu \,A^\nu_{pure} - 
 \partial^\nu \,A^\mu_{pure} - i \,g \,
 [A^\mu_{pure}, A^\nu_{pure}] 
 \ = \ 0 ,  \label{pure-gauge_cond}
\end{eqnarray}
while the second and the third are the
gauge transformation properties for these two components :
\begin{eqnarray}
 A^\mu_{phys}(x) &\rightarrow&
 U(x) \,A^\mu_{phys}(x) \,U^{-1}(x) , \label{homogeneous_tr} \\
 A^\mu_{pure}(x) &\rightarrow&
 U(x) \,\left(\,A^\mu_{pure}(x) - \frac{i}{g}
 \,\, \partial^\mu \,\right) \,U^{-1}(x) . \label{inhomogeneous_tr}
\end{eqnarray}
As a matter of course, these conditions are not enough to fix the decomposition
uniquely. This is not unrelated to the fact that there are many
gauge choices, which nevertheless leads to the same answer for gauge-invariant
observables.\footnote{Remember that a gauge-fixing procedure amounts
to a process of eliminating unphysical gauge degrees of freedom, thereby
selecting out physical degrees of freedom of the gauge field.}
The point of his argument was therefore that one can postpone
a concrete specification of the decomposition until later stage,
while accomplishing a
gauge-invariant decomposition of the QCD angular momentum tensor
$M^{\mu \nu \lambda}$ based on the above conditions only.
As anticipated, he found the existence of two different gauge-invariant
decompositions, which he calls the decomposition (I) and the decomposition (II).
Let us start with the decomposition (II).
The decomposition (II) is a ``seemingly'' covariant generalization of the Chen
decomposition represented as
\begin{eqnarray}
 M^{\mu \nu \lambda}_{QCD} &=& 
 M^{\prime \mu \nu \lambda}_{q-spin} \ + \ 
 M^{\prime \mu \nu \lambda}_{q-OAM} \ + \ 
 M^{\prime \mu \nu \lambda}_{G-spin} \ + \ 
 M^{\prime \mu \nu \lambda}_{G-OAM} \nonumber \\
 &+& \ \mbox{\rm boost} 
 \ + \ 
 \mbox{\rm total divergence} ,
\end{eqnarray}
with
\begin{eqnarray}
 M^{\prime \mu \nu \lambda}_{q-spin}
 &=& \frac{1}{2} \,\epsilon^{\mu \nu \lambda \sigma} \,
 \bar{\psi} \,\gamma_{\sigma} \,\gamma_5 \,\psi , \\
 M^{\prime \mu \nu \lambda}_{q-OAM} 
 &=& \bar{\psi} \,\gamma^{\mu} \,(\,x^{\nu} \,i \,
 D^{\lambda}_{pure}  
 \ - \ x^{\lambda} \,i \,
 D^{\nu}_{pure} \,) \,\psi , \\
 M^{\prime \mu \nu \lambda}_{G-spin}
 &=& 2 \,\mbox{Tr} \,\{\, F^{\mu \lambda} \,A_{phys}^{\nu} \,
 \ - \ F^{\mu \nu} \,A_{phys}^{\lambda} \,\} , \\
 M^{\prime \mu \nu \lambda}_{G-OAM}
 &=& 2 \,\mbox{Tr} \,\{\, F^{\mu \alpha} \,
 (\,x^{\nu} \,D_{pure}^{\lambda}
 \ - \ x^{\lambda} \,D_{pure}^{\nu} \,) \,A_{\alpha}^{phys} \,\} .
\end{eqnarray}
As one sees, the quark and gluon OAMs appearing in this decomposition
are the canonical OAMs aside from the unphysical gauge degrees of
freedom. By this reason, it is sometimes called the ``canonical''
decomposition of the nucleon spin.

On the other hand, the decomposition (I) is a ``seemingly" covariant
generalization of  another noncovariant decomposition due to
Wakamatsu \cite{WakaSD10}. It is given as
\begin{eqnarray}
 M^{\mu \nu \lambda} &=& M^{\mu \nu \lambda}_{q - spin}
 \ + \ M^{\mu \nu \lambda}_{q - OAM} \ + \ M^{\mu \nu \lambda}_{G - spin}
 \ + \ M^{\mu \nu \lambda}_{G - OAM} \nonumber \\
 &+& \ \mbox{\rm boost}  
 \ \ + \ \ \mbox{total divergence} ,
\end{eqnarray}
with
\begin{eqnarray}
 M^{\mu \nu \lambda}_{q - spin}
 &=& M^{\prime \mu \nu \lambda}_{q-spin} , \\
 M^{\mu \nu \lambda}_{q - OAM}
 &=& \bar{\psi} \,\gamma^\mu \,(\,x^{\nu} \,i \,
 D^{\lambda} 
 \ - \ x^{\lambda} \,i \,
 D^{\nu} \,) \,\psi , \\
 M^{\mu \nu \lambda}_{G - spin} 
 &=& M^{\prime \mu \nu \lambda}_{G-spin} , \\
 M^{\mu \nu \lambda}_{G - OAM} 
 &=& M^{\prime \mu \nu \lambda}_{G-OAM}
 \ + \ 2 \,\mbox{Tr} \,[\, (\,D_{\alpha} \,F^{\alpha \mu} \,)
 \,(\,x^{\nu} \,A^{\lambda}_{phys} \ - \ 
 x^{\lambda} \,A^{\nu}_{phys} \,) \,] .
\end{eqnarray}
The quark and gluon OAMs appearing in this decomposition is the mechanical OAMs,
so that it is reasonable to call it the ``mechanical'' decomposition of the
nucleon spin.

One of the greatest advantages of the ``seemingly''
covariant generalization is that it generalizes and unifies the various
nucleon spin decomposition in the market. For example, it was emphasized that
the canonical decomposition (II) reduces to any ones of
Bashinsky-Jaffe \cite{BJ99}, of Chen et al. \cite{Chen08},\cite{Chen09},
and of Jaffe-Manohar \cite{JM90}, after an appropriate
gauge-fixing in a suitable Lorentz frame, which appears to indicate that
they are essentially the same decompositions.\footnote{This statement would
not be correct in the most general context, but it is most probably correct
when applied to the most important longitudinal nucleon spin decomposition,
as we shall discuss throughout this paper.}    
Another advantage of the covariant formulation of the
nucleon spin decomposition is that, it becomes easier to establish explicit 
relations with high-energy DIS observables. In fact,
as pointed out in Ref.~\refcite{WakaSD11A}, the Bashinsky-Jaffe
decomposition obtained based on the light-cone gauge is a special case
of this general treatment.
In Ref.~\refcite{WakaSD11A}, this fact was utilized to show
that the quark and gluon intrinsic spin parts
of the above covariant decomposition precisely coincides with
the first moments of the polarized distribution functions appearing in the
polarized DIS cross-sections.
\begin{eqnarray}
 \Delta q \ = \ \int \,\Delta q(x) \,dx, 
 \ \ \ \ 
 \Delta g \ = \ \int \,\Delta g(x) \,dx .
\end{eqnarray}
Furthermore, based on the mechanical decomposition (I), it can be shown that
the following important relations hold \cite{WakaSD11A} : 
\begin{eqnarray}
 L_q  &=& \langle p \uparrow | \,
 M^{012}_{q-OAM} \,| \,
 p \uparrow \rangle \,/\, \langle p \uparrow |\, p \uparrow \rangle \nonumber \\
 &=& \frac{1}{2} \,\int \,x \,
 [\,H^q (x,0,0) \ + E^q (x,0,0) \,] \,dx \ - \ 
 \frac{1}{2} \,\int \,\Delta q(x) \,dx , \label{Lq1}
\end{eqnarray}
with
\begin{eqnarray}
 M^{012}_{q-OAM} 
 \ = \ \bar{\psi} \,\left(\mbox{\boldmath $x$} \times \frac{1}{i} \,
 \mbox{\boldmath $D$} \right)^3 \,\psi , \label{Lq2}
\end{eqnarray}
and also
\begin{eqnarray}
 L_G  &=& \langle p \uparrow \,| \,
 M^{012}_{G-OAM} \,| \,
 p \uparrow \rangle  \,/\, \langle p \uparrow \,|\, p \uparrow \rangle \nonumber \\
 &=& \frac{1}{2} \,\int \,x \,[\,
 H^G (x,0,0) \ + \ E^G (x,0,0) \,] \,dx \ - \ 
 \int \,\Delta g(x) \,dx , \label{LG1}
\end{eqnarray}
with
\begin{eqnarray}
 M^{012}_{G-OAM} &=& 
 2 \,\mbox{\rm Tr} \,[\,E^j \,(\mbox{\boldmath $x$} \times 
 \mbox{\boldmath $D$}_{pure})^3 \,A^{phys}_j \,] 
 \ + \ 
 2 \,\mbox{\rm Tr} \,[\,\rho \,(\mbox{\boldmath $x$} \times
 \mbox{\boldmath $A$}_{phys})^3 \,] . \label{LG2}
\end{eqnarray}
The relation (\ref{Lq1}) with (\ref{Lq2}) means that the quantity defined as
the difference between the second
moments of unpolarized GPDs $H + E$ and the first moment of polarized
quark distribution just coincides with the proton matrix element of
the quark OAM operator containing full gauge covariant
derivative \cite{Ji97PRL}. 
This just confirms Ji's observation that the quark OAM extracted from the combined
analysis of GPDs and polarized PDFs is the {\it mechanical} OAM not the
{\it canonical} OAM ! The equality (\ref{LG1}) with (\ref{LG2}) provides us with
new information.
It tells that the gluon OAM extracted from the combined analysis of GPD and
polarized PDF contains the {\it potential} OAM, in addition to the gluon
{\it canonical} OAM.
We have pointed out before that the sum of the gluon intrinsic spin and
the gluon OAM in the Jaffe-Manohar decomposition does not coincide with
the gluon total angular momentum in the Ji decomposition. (See Eq.(4).)
The reason of this observation is self-evident now. It is just the
{\it potential angular momentum} that compensates this discrepancy
in such a way that
\begin{equation}
 \Delta G \ + \ L^\prime_G \ = \ J_G \ - \ L_{pot}.
\end{equation}

\section{On the Gauge-Invariant-Extension Approach}

In the previous section, it was pointed out that the three conditions
(\ref{pure-gauge_cond}), (\ref{homogeneous_tr}) and (\ref{inhomogeneous_tr})
are not sufficient to uniquely fix the ``seemingly''
covariant decomposition of the gauge field, $A^\mu (x) = A^\mu_{phys} (x)
+ A^\mu_{pure} (x)$. This is only natural, because any physical condition
necessary for fixing the physical component of $A^\mu$ has not been
imposed at this stage.
Nevertheless, the viewpoint advocated in \cite{WakaSD11A} is that,
since each term of
the decomposition (I) and (II) are clearly gauge-invariant,
one can make any desired gauge choice at the later stage as the needs arise.    
However, quite a different view has rapidly spread out around the
community \cite{JXY12A,JXY12B,Lorce13A,Lorce13B,Lorce13C,Lorce13D,Lorce14,Leader_Lorce13}.
According to this viewpoint, the Chen decomposition is
a gauge-invariant-extension (GIE) of the Jaffe-Manohar decomposition based
on the Coulomb gauge, while the
Bashinsky-Jaffe decomposition (or the Hatta decomposition) is another GIE
of the Jaffe-Manohar decomposition based on the light-cone gauge.
The claim is that, since they are different GIEs, there is no reason that
they give the same physical predictions.

\begin{figure}[h]
\begin{center}
\includegraphics[width=7.0cm]{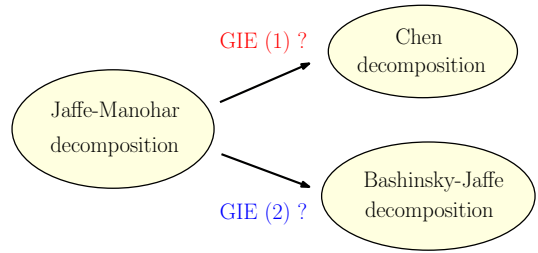}
\caption{\label{Fig2}A schematic picture of the idea of gauge-invariant extension.}
\end{center}
\end{figure}

However, the strangeness of the idea of GIE seems already clear from
the following simple consideration of conceptual nature.
Suppose that the Chen decomposition and the Bashinsky-Jaffe decomposition
(or the Hatta decomposition) are two physically inequivalent GIEs of
the Jaffe-Manohar decomposition. (See Fig.\ref{Fig2}.)
Then, one would immediately encounter the following conceptual questions.
\begin{itemlist}
\item What are the {\it physical meaning} of extended gauge symmetries ?
\item Are there {\it plural} color gauge symmetries in nature ?
\end{itemlist}
Since QCD is a theory with color gauge-invariance from the start, it is
obvious that there is no need of introducing the idea like GIE.
In fact, at least in the simpler case of QED, it was explicitly shown
in Ref.~\refcite{WakaSD12} that the Chen decomposition is {\it never} a
GIE. 
It simply utilizes the gauge degrees of freedom which are present from the start
in the original theoretical expression of the total angular momentum of a coupled
system of the photon and the charged particles.

In the history of the nucleon spin decomposition problem, the word gGIEh was
first introduced by Hoodbhoy and Ji in Ref.~\refcite{Hoodbhoy_Ji99}.
It was used to explain that
the ``should-be'' observable gluon spin $\Delta G$ can be thought of as
agGIEh of the Jaffe-Manohar gluon spin, which is not manifestly gauge-invariant.
The idea was revised in more recent paper by Ji, Xu, and Zhao \cite{JXZ12}.
A similar idea of GIE has been pursued further by Lorc\'{e} based on the
geometrical formulation of gauge
theories \cite{Lorce12,Lorce13A,Lorce13B,Lorce13C,Lorce13D,Lorce14}.
According to Lorc\'{e}, due to the
existence of hidden symmetry called the St\"{u}ckelberg symmetry,
there are in principle infinitely many decompositions of the nucleon spin.

Before explaining his idea in more detail, we point out that the word
``St\"{u}ckelberg symmetry'' is a little disturbing.
The original motivation of St\"{u}ckelberg's idea was to show that
some non-gauge theory can be made a gauge theory by using the so-called
St\"{u}ckelberg trick \cite{Stueckelberg38,Kamefuchi_Umezawa64,Kunimasa_Goto67}. 
The simplest example is provided by the theory
of massive neutral vector boson, the lagrangian of which is given by
\begin{equation}
 {\cal L} \ = \ - \,\frac{1}{4} \,
 (\partial_\mu \,A_\nu - \partial_\nu \,A_\mu)^2 \ - \ 
 \frac{1}{2} \,m^2 \,A_\mu^2 .
\end{equation}
The mass term in this lagrangian obviously breaks the invariance under the gauge
transformation,
\begin{equation}
 A_\mu (x) \ \rightarrow \ A_\mu (x) \ + \ \partial_\mu \Lambda (x) .
\end{equation}
The ``St\"{u}ckelberg trickh begins with introducing an auxiliary
scalar field $\phi (x)$ called the ``compensator'' {\it  by hand},
\begin{equation}
 {\cal L} \ \rightarrow \ {\cal L}^\prime \ = \ 
 - \,\frac{1}{4} \,(\partial_\mu A_\nu - \partial_\nu \,A_\mu)^2
 \ - \ \frac{1}{2} \,m^2 \,(A_\mu - 
 \partial_\mu \phi)^2 .
\end{equation}
If $\phi (x)$ transforms as follows under the gauge transformation, 
\begin{equation}
 \phi (x) \ \rightarrow \ \phi (x) \ + \ \Lambda (x) ,
\end{equation}
then, the new lagrangian ${\cal L}^\prime$ is clearly gauge-invariant !
In this theoretical framework, the original lagrangian ${\cal L}$ is thought to be
a {\it gauge-fixed form} ($\phi = 0$) of the new gauge-invariant lagrangian
${\cal L}^\prime$.
It is clear that this trick enables us to make wide class of nongauge theories
into theories with gauge degrees of freedom, so that it makes the
distinction between true gauge theories and the artificially constructed
gauge theories obscure.

Now let us come back to Lorc\'{e}'s argument. For clarity, we discuss for a
while simpler case of abelian gauge theory. According to him, starting from
some decomposition of the gauge field $A^\mu (x) = A^\mu_{phys} (x) + 
A^\mu_{pure} (x)$, one can get another perfectly acceptable decomposition
$A^\mu (x) = \bar{A}^\mu_{phys} (x) + \bar{A}^\mu_{pure}(x)$, where
\begin{eqnarray}
 \bar{A}^\mu_{phys} (x) &=& A^\mu_{phys} (x) \ - \ \partial^\mu \,C(x), \\
 \bar{A}^\mu_{pure} (x) &=& A^\mu_{pure} (x) \ + \ \partial^\mu \,C(x),
\end{eqnarray}
with $C(x)$ being an arbitrary function of space and time. This then means that
there are infinitely many decompositions of the gauge field into the physical and
pure-gauge components.
It is certainly true that this transformation preserves the pure-gauge-condition
of the pure-gauge component. i.e.
\begin{equation}
 \bar{F}^{\mu \nu} (x) \ = \ \partial^\mu \bar{A}^\nu_{pure} (x) 
 \ - \ \partial^\nu \,\bar{A}^\mu_{pure} (x) \ = \ 0,
\end{equation}
and also its transformation property
\begin{equation}
 \bar{A}^\mu_{pure} (x) \ \rightarrow \ \bar{A}^\mu_{pure} (x) 
 \ + \ \partial^\mu \Lambda (x) ,
\end{equation}
under general gauge transformation
$A^\mu (x) \rightarrow A^\mu (x) + \partial^\mu \,\Lambda (x)$.
However, as emphasized in Refs.~\refcite{WakaSD12} and
\refcite{WakaSD13},
this argument does not pay enough attention to
the fact why $A^{phys}_\mu (x)$ is named the physical component.
As repeatedly emphasized, in the QED case with the noncovariant treatment,
the decomposition by Chen et al. is nothing but the standard decomposition
of the vector potential $\bm{A}$ into the transverse and longitudinal
components as
\begin{equation}
 \bm{A} (x) \ = \ \bm{A}_\perp (x) \ + \ \bm{A}_\parallel (x) ,
\end{equation}
where the two components are respectively required to satisfy the
divergence-free and irrotational conditions : 
\begin{equation}
 \nabla \cdot \bm{A}_\perp \ = \ 0, \ \ \ 
 \nabla \times \bm{A}_\parallel \ = \ 0.
\end{equation}
It is easy to verify that these two components transform as
\begin{eqnarray}
 \bm{A}_\perp (x) &\rightarrow& \bm{A}^\prime_\perp (x) \ = \ 
 \bm{A}_\perp (x), \\
 \bm{A}_\parallel (x) &\rightarrow& \bm{A}^\prime_\parallel (x) \ = \ 
 \bm{A}_\parallel (x) \ - \ \nabla \Lambda (x) ,
\end{eqnarray}
under general gauge transformations. This show that, while $\bm{A}_\parallel$
is a totally {\it arbitrary} quantity that can be changed freely by gauge
transformation, $\bm{A}_\perp$ is essentially a unique object with
definite physical entity.
In fact, within the above noncovariant framework,
the St\"{u}ckelberg transformation {\it a la} Lorc\'{e} reduces
to \cite{Lorce13A,Lorce13B}
\begin{eqnarray}
 \bm{A}_\perp (x) &\rightarrow& \bar{\bm{A}}_\perp (x) \ = \ 
 \bm{A}_\perp (x) \ + \ \nabla C(x), \\
 \bm{A}_\parallel (x) &\rightarrow& \bar{\bm{A}}_\parallel (x) \ = \ 
 \bm{A}_\parallel (x) \ - \ \nabla C(x) .
\end{eqnarray}
One can see that the transformed longitudinal component
$\bar{\bm{A}}_\parallel (x)$ retains the irrotational property,
\begin{eqnarray}
 \nabla \times \bar{\bm{A}}_\parallel \ = \ \nabla \times 
 (\bm{A}_\parallel \ - \ \nabla C(x)) \ = \ \nabla \times \bm{A}_\parallel \ = \ 0.
\end{eqnarray}
(This is simply a reflection of the fact that the standard gauge transformation
for $\bm{A}_\parallel$ keeps the magnetic field
$\bm{B} = \nabla \times \bm{A}$ intact.) In contrast, one finds that the
transformed component $\bar{\bm{A}}_\perp$ does not satisfy the desired
divergence-free (or {\it transversality}) condition
$\nabla \cdot \bar{\bm{A}}_\perp = 0$ any more, since
\begin{eqnarray}
 \nabla \cdot \bar{\bm{A}}_\perp (x) \ = \ \nabla \cdot (\bm{A}_\perp (x) 
 \ + \ \nabla C(x) ) \ = \ \Delta C(x) \ \neq \ 0,
\end{eqnarray}
unless $\Delta C(x) = 0$. The condition $\Delta C(x) = 0$ means that
$C(x)$ is a harmonic function in three spatial dimension. 
If it is required to vanish at the spatial infinity, it must be
identically zero, i.e. $C(x) \equiv 0$. 
This means that the invariance under the St\"{u}ckelberg transformation
actually does not exist, provided that an appropriate physical condition,
i.e. the transversality condition, is imposed on the physical component
of the vector potential $\bm{A}$. To be more strict, there are some
matters to be attended.
The boundary condition $\lim_{|\bm{x}| \rightarrow \infty} \,C(x) = 0$
may not be always satisfied in some singular gauges like the light-cone gauge.  
More importantly, the notion of transversality depends on
the Lorentz frame of reference. Namely, a vector field that appears transverse
may not necessarily be transverse in another Lorentz frame.
This indicates that, if the St\"{u}ckelberg symmetry as proposed by Lorc\'{e}
{\it does} exist, one of its origin would be the Lorentz-frame dependence
of the decomposition $A^\mu (x) = A^\mu_{phys} (x) + A^\mu_{pure} (x)$.

In a recent paper \cite{Lorce14}, Lorc\'{e} argued about another possible
origins of the St\"{u}ckelberg symmetry. 
One is the path-dependence of the decomposition
$A^\mu (x) = A^\mu_{phys} (x) + A^\mu_{pure} (x)$, which arises within
the formulation using the gauge link or the Wilson line. (In a recent paper,
Tiwari discussed the role of topology in the path-dependence \cite{Tiwari13}.)
The other is the background dependence of the decomposition, which arises
in the formulation based on the background field method of gauge theories.
In the following, we discuss only the former, because it is more closely
connected with actual physical situations which we encounter in the studies of
high-energy DIS observables. By making use of a path-dependent
Wilson line ${\cal W}_C (x,x_0) = {\cal P} \,\left[ e^{\,i \,g \,\int_{x_0}^x \,
A_\mu (s) \,d s^\mu} \right]$ with $x_0$ being an appropriate reference point,
it is possible to give an explicit form of the decomposition
\begin{equation}
 A^\mu (x) \ = \ A^\mu_{phys} (x) \ + \ A^\mu_{pure} (x) ,
\end{equation}
where
\begin{eqnarray}
 A^\mu_{phys} (x) &=& - \,\int_{x_0}^x \,{\cal W}_C (x,s) \,\,
 F_{\alpha \beta} (s) \,\,{\cal W}_C (s,x) \,\,
 \frac{\partial s^\alpha}{\partial x_\mu} \,\,d s^\beta, \\
 A^\mu_{pure} (x) &=& \ \frac{i}{g} \,\,{\cal W} (x,x_0) \,\,
 \frac{\partial}{\partial x_\mu} \,\,{\cal W}_C (x_0,x) .
\end{eqnarray}
Changing the path $C$ alters both of $A^{phys}_\mu (x)$ and $A^{pure}_\mu (x)$,
but the sum of them is intact. According to Lorc\'{e}, the freedom in the choice
of the path is therefore an origin of the St\"{u}ckelberg symmetry.
Accordingly, St\"{u}ckelberg-invariant quantities must be path-independent,
whereas path-dependent quantities are called St\"{u}ckelberg non-invariant.
Based on these considerations, Lorc\'{e} comes to conclude that one should distinguish
the two forms of gauge-invariance : 

\begin{itemlist}
\item {\it Strong gauge-invariance} : Quantities with this invariance are invariant
under St\"{u}ckelberg transformation as well as under the usual gauge
transformation.
They are local quantities and can be measured
without relying upon any expansion framework like the twist-expansion in
perturbative QCD.

\item {\it Weak gauge-invariance} : Quantities with this invariance are
invariant under the usual gauge transformation, but it is not St\"{u}ckelberg
invariant. They are generally nonlocal. They correspond to 
``quasi-observables'', which can be measured only within an certain
expansion framework. Putting it in another way, they are theoretical-scheme
dependent (quasi-)observables.

\end{itemlist}

All these statements might be correct in the most general context.
Still, what is lacking in this general argument with highly mathematical
nature is the insight into the physical meaning of the decomposition
$A_\mu (x) = A^{phys}_\mu (x) + A^{pure}_\mu (x)$.
We know that, among the four components of the gauge field $A_\mu (x) \,
(\mu = 0,1,2,3)$, independent dynamical degrees of freedom are
only two \cite{BookBD65}.
They are two transverse components, say $A^1$ and $A^2$.
(The other two components, i.e. the so-called scalar component $A^0$
and longitudinal component $A^3$, are not independent dynamical degrees
of freedom.)
Undoubtedly, what should be identified with the {\it physical} components
of $A_\mu (x)$ are these two transverse components.
Unfortunately, as repeatedly emphasized, a delicacy here is the fact that the
notion of transversality depends on the Lorentz frame.
The standard transverse-longitudinal decomposition
$\bm{A} (x) = \bm{A}_\perp (x) + \bm{A}_\parallel (x)$
in the noncovariant framework has a meaning only after specifying
a working Lorentz-frame of reference.
Nevertheless, it is important to recognize the fact that one has a freedom to
start this noncovariant decomposition in an arbitrarily chosen Lorentz frame.
The question is therefore whether a quantity we are discussing
is a Lorentz-frame-dependent quantity or not.
If the quantity of question is a Lorentz-frame-independent one, there is no
reason to suspect that the noncovariant treatment as proposed by Chen et al.
would give a wrong answer.

Returning to our nucleon spin decomposition problem, a possible candidate of
Lorentz-invariant spin observable is helicity.\footnote{The helicity is
exactly Lorentz-invariant only for a
massless particle, but its invariance holds also for a massive particle like
the nucleon for a wide-class of Lorentz transformations, which are relevant
for our later discussion.}
This suggests that the nucleon helicity sum rule or the longitudinal nucleon
spin decomposition can be made so as to meet both the requirements
of gauge-invariance and Lorentz-invariance.
However, this conversely indicates that one cannot expect the same for
more general nucleon spin decomposition like the decomposition of the
transverse nucleon spin.
At any rate, when discussing the possibility of gauge-invariant
decomposition of the nucleon spin, one should pay more attention to the
significant difference between the above two cases.
Undoubtedly, what makes our problem
difficult is an {\it intricate interplay} between the gauge- and Lorentz-frame
dependencies of the nucleon spin decomposition.
 
Before ending this section, we think it useful to explain the reason
why a lot of researchers believe that the Chen decomposition \cite{Chen08,Chen09}
and the Bashinsky-Jaffe decomposition \cite{BJ99} 
(or Hatta decomposition \cite{Hatta11}) are physically
inequivalent GIEs of the Jaffe-Manohar decomposition \cite{JM90}. 
The reason can be traced back to Chen et al.'s calculations for the
evolution matrix of the quark and gluon longitudinal momentum
fraction \cite{Chen09} and also for quark and gluon longitudinal spin
based on their noncovariant decomposition \cite{Chen11B}.
For example, their prediction for the asymptotic value of the gluon
momentum fraction in the nucleon
\begin{equation}
 \lim_{Q^2 \rightarrow \infty} \,\,\langle x \rangle^G  
 \ = \ \frac{8}{8 + 6 \,n_f} \ \ \sim \ \ \frac{1}{5} ,
\end{equation}
is drastically different from the standardly-believed value : 
\begin{equation}
 \lim_{Q^2 \rightarrow \infty} \,\,\langle x \rangle^G  
 \ = \ \frac{16}{16 + 3 \,n_f} \ \ \sim \ \ \frac{1}{2} ,
\end{equation}
which can properly be reproduced in the light-cone gauge calculation.
It was conjectured in Ref.~\refcite{WakaSD12} and \refcite{WakaSD13}
that the calculation of the evolution matrices in the
Coulomb gauge (or in the Chen decomposition) is highly nontrivial,
and that Chen et alfs calculation is probably wrong.
This conjecture was, in a sense, explicitly confirmed by the recent study
by Ji, Zhang, and Zhao \cite{JZZ13}. 
For simplicity, we explain the point in simpler Abelian case (QED).
They start with the definition of gauge-invariant photon spin {\it a la}
Chen et al.
\begin{equation}
 {\bm S}_\gamma \ = \ \bm{E} \times \bm{A}_\perp ,
\end{equation}
where $\bm{A}_\perp$ is the physical or transverse component of the photon
field given as
\begin{equation}
 {\bm A}^i_\perp \ = \ \left( \delta^{ij} \ - \  
 \frac{\nabla_i \,\nabla_j}{\nabla^2} \right) \,{\bm A}^j .
\end{equation}
Now, consider a boost along the negative 3-axis with infinite velocity. 
In this IMF limit, they found that $\bm{S}_\gamma$ takes the following form :
\begin{equation}
 {\bm S}_\gamma \ = \ 
 {\bm E} \times {\bm A}_\perp \ \ \rightarrow \ \  
 {\bm E} \times \left( \bm{A} \ - \  
 \frac{1}{\nabla^+} \,\nabla \,A^+ \right) .
\end{equation}
Here, the quantity
\begin{equation}
 \bm{A}_{phys} \ = \ \bm{A} \ - \  
 \frac{1}{\nabla^+} \,\nabla \,A^+
\end{equation}
is basically the physical component of the photon in the light-cone
gauge.\footnote{It is important to recognize that this $\bm{A}_{phys}$
is nonlocal, but it is nevertheless path-independent. 
These two types of nonlocality should be clearly distinguished.
See the disussion in Ref.~\refcite{WakaSD13}, for more detail.}
A delicacy here is that the IMF limit and the loop-integrals necessary for
the calculation of the evolution matrix are {\it noncommutable}.
They argue that, if one properly takes care of this noncommutativity,
the Chen decomposition gives exactly the same answer as that in the
light-cone gauge.  
This is nice, but there still remains a delicate question.
In their whole analysis, they must eventually take the IMF limit. In this sense,
the IMF limit (and the light-cone gauge) still play uninterchangeable role !
On the other hand, however, one usually believes that the longitudinal gluon spin,
or more generally the longitudinal gluon spin distribution in the nucleon,
is a Lorentz-frame-independent quantity, even though its physical interpretation
becomes simplest in the IMF.
Is $\Delta G$ not only gauge-invariant but also Lorentz-frame independent
observable ? It appears that their analysis has not given a completely
satisfactory answer to this question.

\section{Recent Controversies on Transverse Nucleon Spin Decomposition}

By historical as well as practical reason, our main interest so far has
been devoted to the problem of longitudinal nucleon spin sum rule.
Naturally, one expects to get useful complimentary information
from studies of transverse spin
sum rule, which has in fact been a object of intense debate in a few years.
The importance of comparing the transverse and longitudinal nucleon spin
decomposition in the study of relativistic nucleon spin observables was first
emphasized in the paper by Bakker, Leader and Trueman (BLT) \cite{BLT04}.
They proposed a transverse spin sum rule which contains the contribution
from the quark transversity distributions. However, as criticized by Ji, Xiong,
and Yuan \cite{JXY12B}, since the quark transversity is a chiral-odd object,
the proposed sum rule appears in direct contradiction with the chiral-even
property of the nucleon
spin and orbital angular momentum. Later, Leader proposed another transverse
spin sum rule \cite{Leader12} based on the manipulation in the BLT paper
\cite{BLT04}. 
In the case of a transversely polarized nucleon, moving along the positive
$z$ axis, he obtains the sum rule : 
\begin{equation}
 J^\perp_q \ = \ \frac{1}{2 \,M} \,
 \left[\, P_0 \,\int_{-1}^1 \,x \,E_q (x,0,0) \,dx \ + \ M \,
 \int_{-1}^1 \,x \,H_q (x,0,0) \,dx \, \right] ,
\end{equation}
where $P_0$ is the energy of the nucleon. Undoubtedly, this is a
Lorentz-frame-dependent sum rule.
The reason is simple. This sum rule is obtained
by taking a nucleon matrix element of the angular momentum operator $J^x_q$,
which is not invariant under the Lorentz-boost in the $z$ direction, i.e.
in the direction along which the nucleon is moving.

With the intension of obtaining a boost-invariant transverse spin sum rule for
a moving nucleon (along the $z$ direction), Ji, Xiong, and Yuan attempted to
construct the transverse spin sum rule based on the Lorentz-covariant
Pauli-Lubanski vector \cite{JXY12B,Lubanski42}.
Their sum rule is given as
\begin{equation}
 J^\perp_q \ = \ \frac{1}{2} \,
 \left[\, \int_{-1}^1 \,x \,E_q (x,0,0) \,dx \ + \ 
 \int_{-1}^1 \,x \,H_q (x,0,0) \,dx \, \right] ,
\end{equation}
which just takes the same form as the corresponding longitudinal nucleon spin
sum rule (the well-known Ji sum rule \cite{Ji97PRL}) given as
\begin{equation}
 J^\parallel_q \ = \ \frac{1}{2} \,
 \left[\, \int_{-1}^1 \,x \,E_q (x,0,0) \,dx \ + \ 
 \int_{-1}^1 \,x \,H_q (x,0,0) \,dx \, \right] ,
\end{equation} 
and it appears to be Lorentz-frame-independent.
However, a flaw of their derivation was pointed out by several
authors \cite{LL12,Leader13, HTY13, HKMR13,HKM14}.
Their derivation is based on the parametrization of the
nucleon matrix element of the quark and gluon energy momentum tensor : 
\begin{eqnarray}
 \langle P^\prime S \,|\, T^{\mu \nu}_i (0) \,|\,P S \rangle \nonumber
 &=& \bar{U} (P^\prime) \,
 \left[\,A_i (\Delta^2) \,\gamma^{\{\mu} \,\bar{P}^{\nu\}} \ + \ 
 B_i (\Delta^2) \,\frac{\bar{P}^{\{\mu} \,i \,\sigma^{\nu\}} \,\Delta_\alpha}{2 \,M}
 \right. \nonumber \\
 &\,& \left. + \ 
 C_i (\Delta^2) \,\frac{\Delta^\mu \,\Delta^\nu - g^{\mu \nu}\,\Delta^2}{M}
 \ + \ \bar{C}_i (\Delta^2) \,M \,g^{\mu \nu} \,\right] \,U (P), \ \ \ 
\end{eqnarray}
where $\bar{P} = (P^\prime + P) \,/\,2, \Delta = P^\prime - P$, and
$A_i, B_i, C_i$ and $C^\prime_i$ are generalized form factors, with $i$
denoting either of quark or gluon. In deriving the above sum rule, 
they erroneously dropped the possible contributions from the $\bar{C}_i$ terms.
Later, several authors rederived the transverse nucleon spin sum rule,
by careful account of the $\bar{C}_i$ terms \cite{Leader13,HTY13,HKM14}.
However, a worry is that
their answers turned out to be totally diverging.
For instance, Leader arrived at the answer \cite{Leader13}
\begin{eqnarray}
 J^\perp_{q/G} \ = \ \frac{1}{2} \,
 \left( A_{q/G} (0) + B_{q/G} (0) \right)
 \ + \ \frac{P^0 - M}
 {2 \,P^0} \,\,\bar{C}_{q/G} (0),
\end{eqnarray}
with the relation
\begin{eqnarray*}
 A_{q/G} (0) &=& \int_{-1}^1 \,x \,H_{q/G} (x,0,0) \,dx  , \\
 B_{q/G} (0) &=& \int_{-1}^1 \,x \,E_{q/G} (x,0,0) \,dx .
\end{eqnarray*}
On the other hand, Hatta, Yoshida, and Tanaka (HYT) gave \cite{HTY13}
\begin{eqnarray}
 J^\perp_{q/G} \ = \ \frac{1}{2} \,\,
 \left( A_{q/G} (0) + B_{q/G} (0) \right)
 \ + \ \frac{P^3}
 {2 \,(P^0 + M)} \,\,\bar{C}_{q/G} (0).
\end{eqnarray}
Finally, based on the framework of light-front quantization scheme,
Harrindranath, Kundu, and Mukerjee (HKM) arrived at the following
answer \cite{HKM14},  
\begin{eqnarray}
 J^\perp_{q/G} \ = \ \frac{1}{2} \,
 \left( A_{q/G} (0) + B_{q/G} (0) + \bar{C}_{q/G} (0) \right).
\end{eqnarray}
Several comments are in order here. Because of the relation,
\begin{equation}
 \bar{C}_q (0) \ + \ \bar{C}_G (0) \ = \ 0,
\end{equation}
all these three results are consistent with the net transverse nucleon
spin sum rule : 
\begin{equation}
 J^\perp_q \ + \ J^\perp_G \ = \ \frac{1}{2} \,
 \left[\,A_q (0) \ + \ B_q (0) \ + \ A_G (0) \ + \ B_G (0) \,\right]
 \ = \ \frac{1}{2}.  
\end{equation}
One also observes that all these three expressions coincide in the
infinite-momentum-frame (IMF) limit
$P^2 \rightarrow \infty$, $P^0 \rightarrow \infty$.
Note, however, that the first two sum rules are generally dependent
on the nucleon momentum or energy, so that they are obviously
Lorentz-frame-dependent. 

\vspace{2mm}
\renewcommand{\arraystretch}{1.6}
\begin{table}[h]
\tbl{The various choices of angular momentum tensor and nucleon spinors
for obtaining transverse spin sum rule.}
{\begin{tabular}{@{}ccc@{}} 
\hline
 & $J^{\alpha \beta}$ & $|\,P S^x \rangle$ \\
\hline
Leader & $\int \,d^3 x \,M^{0 \alpha \beta}$ & 
Dirac spinors \\
\hline
HTY & $\int \,d x^- \,d^2 x^\perp \,
M^{+ \alpha \beta}$ & 
Dirac spinors \\
\hline
\ \ HKM \ \ & \ \ $\int \,d x^- \,d^2 x^\perp \,
M^{+ \alpha \beta}$ \ \ 
& \ \ Light-front spinors \ \ \\
\hline
\end{tabular} \label{transSR}}
\end{table}

How can we understand these differences ?
The reason is that they all calculated the nucleon matrix element
of the Pauli-Lubanski vector
$W^x$ between the transversely polarized nucleon state in the $x$ direction : 
\begin{eqnarray}
 \langle P S^x \,| \, W^x \,|\,P S^x \rangle \ \ \ 
 \mbox{\tt with} \ \ 
 W_\mu \ = \ - \,\frac{1}{2} \,
 \epsilon_{\mu \alpha \beta \rho} \,J^{\alpha \beta} \,P^\rho ,
\end{eqnarray}
but with different angular momentum tensor $J^{\alpha \beta}$ and
different nucleon spinors.
As summarized in Table \ref{transSR}, Leader uses the angular momentum tensor
$M^{0 \alpha \beta}$ in the equal-time (ET) formalism together with Dirac spinors.
Hatta et al. (HTY) use the angular momentum tensor in the light-front (LF) formalism
$M^{+ \alpha \beta}$ together with Dirac Spinors.
On the other hand, Harindranath et al. (HKM) use the angular momentum
tensor in the LF formalism together with the light-front spinors.
HKM emphasize that their result based on the LF (light-front) formalism is absolutely Lorentz-frame-independent, but this statement is misleading.
It is known that the use of the LF spinors in the LF formalism is equivalent to
working in the IMF.
In the IMF, however, the dependence on the nucleon longitudinal momentum $P^3$ is naturally washed out.
What HKM have shown is actually the $P_\perp$-independence of their sum
rule \cite{HKM14}.


In any case, one now convinces that the transverse spin sum rule
is generally
Lorentz-frame-dependent due to the existence of the term $\bar{C}(0)$.
It is very important to recognize the fact that the existence of plural forms of transverse spin decomposition has nothing to do with our {\it gauge problem},
because both of $J_q$   and $J_G$ are obviously gauge-invariant.
Rather, one can say that the origin of existence of plural forms of transverse
spin sum rule is the {\it relativity} !

In fact, as nicely reviewed in Ref.\refcite{PGW13}, the treatment of spin
in relativistic quantum mechanics is far more complicated than in
nonrelativistic quantum mechanics.
The relevant complication is known to arise from the fact that
the sequences of rotationless Lorentz boosts can
generate rotations as dictated by the following commutation relation : 
\begin{equation}
 [K_i, K_j] \ = \ - \,i \,\epsilon_{ijk} \,J_k ,
\end{equation}
where $K_i \,(i = 1,2,3)$ are the rotationless boost generators and
$J_i \,(i = 1,2,3)$ are the spatial rotation generators.
For a particle with nonzero mass, it is most convenient to define its
spin states in its rest frame\footnote{The treatment of massless particles
requires another care, since there is no rest frame for a massless
particle.}. However, the relativistic spin states
(or observables) generally depend both on the frame where the spin is
defined and on a set of Lorentz transformations which relates the
frame where the spin is defined and a frame where a particle has a
definite momentum $\bm{p}$. This in principle induces an infinite
numbers of possible choices of spin observables in relativistic
quantum mechanics.

Despite the above general statement, it is very important to recognize
a remarkable difference between the transverse spin decomposition and the
longitudinal one.
In fact, one can easily verify that any of the above-mentioned three choices
for $J^{\alpha \beta}$ and the nucleon spinors leads
to exactly the same sum rule for the longitudinal nucleon spin,
\begin{eqnarray}
 J^\parallel_{q/G} \ = \ \frac{1}{2} \,
 \left(\,A_{q/G} (0) \ + \ B_{q/G} (0) \,\right)
\end{eqnarray}
which is nothing but the celebrated Ji sum rule.
Important lessons learned from these observations are as follows.
First, since the Lorentz-frame-independent decomposition of the transverse
nucleon spin into the quark and gluon total angular momenta seems to
be impossible, we naturally have no chance to get further
frame-independent decomposition of
the quark and gluon transverse angular momenta into their intrinsic spin and
orbital part. In contrast, we still have a possibility to get a gauge- and
frame-independent decomposition of the longitudinal $J_q$ and $J_G$ into
their intrinsic spin and orbital parts.
This is true even for the massless gluon, for which
there is no rest frame.
In Sect.5, we shall demonstrate how it is possible for easier
QED case, i.e. for massless photons.

\section{Does the ``Canonical Orbital Angular Momentum'' Satisfy the SU(2)
Commutation Relation ?}

Quite a lot of people believe that a greatest advantage of the canonical type
decomposition of the nucleon spin is that each term satisfies the angular momentum
commutation relation, and that this is of vital
importance for natural interpretation of each term as an angular momentum. 
One can convince below that this is not necessarily true even in the simpler
case of Abelian gauge theory.
(Here we closely follow the argument given in Ref.~\refcite{VanEN94A}
and \refcite{VanEN94B}. 
See also the standard textbooks \cite{BookCDG89,BookJR76,BookBLP82}
for general discussion.)
The fact is that the observability of the photon spin and OAM
has little to do with their SU(2) commutation relations. 

Let us start with the textbook expression for the total photon angular momentum,
\begin{eqnarray}
 \bm{J} &=& \int \,\bm{r} \times
 (\bm{E} \times \bm{B}) \,\, d^3 r.
\end{eqnarray}
There is no doubt that this expression is manifestly gauge-invariant.
As is well-known, after an appropriate choice of the Lorentz-frame, the vector
potential $\bm{A}$ of the photon can be gauge-invariantly decomposed into the
transverse and longitudinal parts as
\begin{eqnarray}
 \bm{A} (x) \ = \ \bm{A}_\perp (x) \ + \ \bm{A}_\parallel (x)
 \ \equiv \ 
 \bm{A}_{phys} (x) \ + \ \bm{A}_{pure} (x).
\end{eqnarray}
This gives the corresponding transverse-longitudinal decomposition of the
electric field,
\begin{eqnarray}
 \bm{E} \ = \ \bm{E}_\perp \ + \ \bm{E}_\parallel,
\end{eqnarray}
with 
\begin{eqnarray}
 {\bm E}_\perp &=& - \,\frac{\partial {\bm A}_\perp}{\partial t},
 \ \ \ 
 {\bm E}_\parallel \ = \ - \,\nabla A^0 \ - \ 
 \frac{\partial {\bm A}_\parallel}{\partial t} .
\end{eqnarray}
Correspondingly, the total photon angular momentum can be decomposed
into two parts as
\begin{eqnarray}
 \bm{J} &=& \int \,\bm{r} \times 
 (\bm{E}_\parallel \times \bm{B}) \,d^3 r \ + \ 
 \int \,\bm{r} \times 
 (\bm{E}_\perp \times \bm{B}) \,d^3 r \\
 &\equiv& \ \hspace{10mm} \bm{J}_{long}
 \hspace{10mm} + \hspace{10mm} \bm{J}_{trans} .
\end{eqnarray}
By using the Gauss law $\nabla \cdot {\bm E}_\parallel \ = \ \rho$,
it is easy to show that the longitudinal part $\bm{J}_{long}$, which contains
the $\bm{E}_\parallel$ component, can be rewritten in the form : 
\begin{eqnarray}
 \bm{J}_{long}
 &=& \int \,\rho \,
 (\bm{r} \times \bm{A}_\perp) \,d^3 r \ \equiv \ \bm{L}_{pot},
\end{eqnarray}
which is nothing but the ``potential angular momentum'' in the terminology
of Ref.~\refcite{WakaSD10}. On the other hand, the transverse part
$\bm{J}_\perp$ can further be decomposed into two parts as,
\begin{eqnarray}
 \bm{J}_{trans}
 &=& \int \,E^l_\perp \,
 (\bm{r} \times \nabla) \,A^l_\perp \,d^3 r \ + \ 
 \int \,\bm{E}_\perp \times \bm{A}_\perp \,d^3 r
 \ = \ {\bm L} \ + \ 
 {\bm S} .
\end{eqnarray}
which can be identified with the ``canonical'' OAM
and the intrinsic spin of the photon. We emphasize that this
decomposition is gauge-invariant, because $\bm{A}_\perp$ is gauge-invariant
($\bm{E}_\perp$ is naturally gauge-invariant).
We also recall the fact that the sum of the potential angular
momentum $\bm{J}_{long} = \bm{L}_{pot}$ and the ``canonical'' OAM
$\bm{L}^\gamma_{can} \equiv \int \,d^3 r \,E^\perp_l \,(\bm{r} \times \nabla)
\,A^\perp_l$ can be identified with the ``mechanical'' OAM
$\bm{L}^\gamma_{mech}$ of the {\it photon}.
Note, however, that, $\bm{L}_{pot} = 0$ for free
photons (since $\rho = 0$), so that in this case, there is no difference
between the ``mechanical'' and ``canonical'' OAMs.
This is the situation which we shall consider below, since
very weak interactions between photons and charged particles is allowed to be
introduced only at the final stage as as a perturbation.
 
To proceed, we first introduce transverse mode functions with polarization
$\lambda$ as solutions of Helmholtz equation (it is nothing but the
Maxwell equation for a free photon) with the transversality condition :
\begin{equation}
 \nabla^2 \,\bm{F}_\lambda \ = \ - \,k^2 \,\bm{F}_\lambda,
 \ \ \ \nabla \cdot \bm{F}_\lambda \ = \ 0. \\
\end{equation}
They are supposed to satisfy the following orthnormalization condition : 
\begin{eqnarray}
 \langle \bm{F}_\lambda \,|\, \bm{F}_{\lambda^\prime} \rangle
 \ \equiv \ \int \,\bm{F}_\lambda \cdot \bm{F}_{\lambda^\prime} \,d^3 r
 \ = \ \delta_{\lambda \lambda^\prime} .
\end{eqnarray}
The simplest choice for the transverse mode functions would be the circularly
polarized plane waves :
\begin{eqnarray}
 \bm{F}_\lambda \ = \ \frac{1}{\sqrt{V}} \,\,
 \bm{\varepsilon}_{\bm{k},s} \,\,e^{i \,\bm{k} \cdot \bm{r}}
 \ \ \ (s \,= \, \pm \,1) ,
\end{eqnarray}
but we can also take other choices like that of the para-axial laser beams
used in the measurements of the photon orbital angular
momentum \cite{Beth36,ABSW92,VanEN92}.
With these mode functions, the electromagnetic field or vector
potential can be expanded as
\begin{eqnarray}
 \bm{A}_\perp &=& \sum_\lambda \,\sqrt{\frac{\hbar}{2 \,\omega_\lambda}}
 \,\,\left[\,a_\lambda \,\bm{F}_\lambda \ + \ 
 a_\lambda^\dagger \,\bm{F}_\lambda^* \,\right],
\end{eqnarray}
where $a_\lambda$ and $a^\dagger_\lambda$ are the annihilation and creation
operator of the photon with the polarization $\lambda$, satisfying the
standard commutation relation :
\begin{eqnarray}
 [ a_\lambda, a_{\lambda^\prime}^\dagger ] \ = \ 
 \delta_{\lambda \lambda^\prime} .
\end{eqnarray}
The corresponding mode-expansions for the electric and magnetic fields are
then given by
\begin{eqnarray}
 \bm{E}_\perp &=& \sum_\lambda \,i \,
 \sqrt{\frac{\hbar \,\omega_\lambda}{2}}
 \,\,\left[\,a_\lambda \,\bm{F}_\lambda \ - \ 
 a_\lambda^\dagger \,\bm{F}_\lambda^* \,\right] , \\
 \bm{B}_\perp &=& \sum_\lambda \,i \,
 \sqrt{\frac{\hbar}{2 \,\omega_\lambda}}
 \,\,\left[\,a_\lambda \,\nabla \times \bm{F}_\lambda \ + \ 
 a_\lambda^\dagger \,\nabla \times \bm{F}_\lambda^* \,\right].
\end{eqnarray}
Using these formulas, we are led to the second-quantized forms of the
intrinsic spin and ``canonical'' OAM of the photon as
\begin{eqnarray}
 \bm{S}
 &\equiv& \int \,\bm{E}_\perp \times \bm{A}_\perp \,d^3 r
 \ = \ \frac{1}{2} \,\sum_{\lambda, \lambda^\prime} \,\,
 [ a_\lambda^\dagger \,a_{\lambda^\prime} \ + \ 
 a_{\lambda^\prime} \,a_\lambda^\dagger ] \,\,
 \langle \bm{F}_\lambda \,| \,
 \hat{\bm{S}} \,|\, 
 \bm{F}_{\lambda^\prime} \rangle , \\
 \bm{L}  
 &\equiv& \int \,\bm{E}^l_\perp 
 \,(\bm{r} \times \nabla) \,\bm{A}^l_\perp \,d^3 r
 \ = \ \frac{1}{2} \,\sum_{\lambda, \lambda^\prime} \,\,
 [ a_\lambda^\dagger \,a_{\lambda^\prime} \ + \ 
 a_{\lambda^\prime} \,a_\lambda^\dagger ] \,\,
 \langle \bm{F}_\lambda \,| \,
 \hat{\bm{L}} \,|\, 
 \bm{F}_{\lambda^\prime} \rangle ,
\end{eqnarray}
with the definitions of the operators $\hat{\bm{S}}$ and $\hat{\bm{L}}$ :
\begin{eqnarray}
 (\hat{\bm{S}})_{ij} 
 \ = \ - \,i \,\hbar \,\varepsilon_{ijk}, \ \ \ 
 \hat{\bm{L}} 
 \ = \ - \,i \,\hbar \,(\bm{r} \times \nabla)
\end{eqnarray}
These operators $\hat{\bm{L}}$ and $\hat{\bm{S}}$ certainly satisfy the familiar
SU(2) algebra :
\begin{eqnarray}
 [\,\hat{S}_i, \hat{S}_j \,] \ = \ i \,\hbar \,\hat{S}_k, \ \ \ 
 [\,\hat{L}_i, \hat{L}_j \,] \ = \ i \,\hbar \,\hat{L}_k .
\end{eqnarray}
However, the crucial point here is that what correspond to observables are
not $\hat{\bm{S}}$ and $\hat{\bm{L}}$ but $\bm{S}$ and $\bm{L}$,
because the latter are operators acting on {\it physical Fock space}.
What are the commutation relations of $\bm{S}$ and $\bm{L}$ like, then ?
To find them, 
choose circularly polarized plane waves again as field modes,
\begin{eqnarray}
 \bm{F}_\lambda \ = \ \frac{1}{\sqrt{V}} \,\,
 \bm{\varepsilon}_{\bm{k},s} \,\,e^{i \,\bm{k} \cdot \bm{r}}
 \ \ \ (s \,= \, \pm \,1) .
\end{eqnarray}
In this case, $\bm{S}$ is represented as
\begin{eqnarray}
 \bm{S} \ = \ \sum_{\bm{k}} \,
 \frac{\hbar \,\bm{k}}{k} \,(\,a_{\bm{k},1}^\dagger \,a_{\bm{k},1} \ - \ 
 a_{\bm{k},-1}^\dagger \,a_{\bm{k},-1} \,) .
\end{eqnarray}
Note that $\bm{S}$ is given as a superposition of the number operators of photons
with definite polarizations, so that any components of $\bm{S}$ must commute :
\begin{eqnarray}
 [ S_i, S_j ] \ = \ 0 .
\end{eqnarray}
Somewhat unexpectedly, we therefore find that $\bm{S}$ does not satisfy
standard angular momentum commutation relation.
This means that $\bm{S}$ does not generate general rotations of photon
polarization states. 
Instead, it generates a transformation of the polarization vector such that
the transversality of the polarization vector is preserved.
To be more concrete, under a rotation by angle $\theta$ about the direction
of the photon momentum $\bm{k}$, the photon state with helicity $\lambda$
transforms as \cite{BookLeader01}
\begin{equation}
 |\,\bm{k} \,; \,\lambda \rangle \ \rightarrow \ 
 \exp \,\left[\, - \,i \,\theta \,\frac{\bm{S} \cdot \bm{k}}{|\bm{k}|} \,\right]
 \,| \,\bm{k} \,;\, \lambda \rangle \ = \ e^{\,- \,i \,\theta \,\lambda} \,
 |\,\bm{k} \,;\,\lambda \rangle .
\end{equation}
This shows that only the components of the operator $\bm{S}$ along
$\bm{k}$. i.e. the (the propagation direction of the photon) is a true
spin angular momentum operator in the sense that this component certainly
generate spin rotation. The components of the operator $\bm{S}$ along $\bm{k}$ is
nothing but the {\it helicity} of the photon.

What about the commutation relation of $\bm{L}$, then ?
First, notice that the total photon angular momentum must satisfy the
standard commutation relation,
\begin{eqnarray}
  [J_i, J_j] \ = \ i \,\hbar \,\varepsilon_{ijk} \,J_k .
\end{eqnarray}
Second, $\bm{S}$ and $\bm{L}$ must transform as vectors under spatial rotation,
so that
\begin{eqnarray}
 \,[ J_i, S_j ] &=& i \,\hbar \,\varepsilon_{ijk} \,S_k \\
 \,[ J_i, L_j ] &=& i \,\hbar \,\varepsilon_{ijk} \,L_k .
\end{eqnarray}
Combining these relations with the commutation relation $[S_i, S_j] = 0$
for $\bm{S}$, it follows that
\begin{eqnarray}
 [ L_i, L_j ] 
 &=& i \,\hbar \,\varepsilon_{ijk} \,
 (L_k \,- \,S_k) \\
 \,[ L_i, S_j ] &=& i \,\hbar \,\varepsilon_{ijk} \,S_k 
\end{eqnarray}
Thus, one clearly sees that $\bm{L}$ does not satisfy the standard angular
momentum algebra either, even though it is the very quantity, which can
be measured, for instance, as OAM of para-axial
laser beam \cite{Beth36,ABSW92,VanEN92}.
We therefore confirm that, for massless particles like
photons (and naturally also for gluons), there is no connection between the
observability and the requirement of the SU(2) commutation relation
of each piece of spin and orbital angular momentum decomposition.
All these delicacies of photon (or gluon) spin decomposition comes from the
fact that there is no rest frame for massless particles!
In a mathematical language, as emphasized by Zhang and Pak \cite{ZhangPak12},
the only frame-independent notion of spin for a massless particle is the
{\it helicity}, which can be described by a little group $E(2)$ of
the Lorentz group.

Here we do not go further into the detail. But, by introducing the interaction
of the photon beam with atoms, the following conclusion can be drawn.
(Interested readers are recommended to consult with the original
papers \cite{VanEN94A,VanEN94B}.) 
Both gspinh$\bm{S}$ and gorbitalh angular momentum $\bm{L}$ of a photon are
well defined quantities and might in principle separately be measured. 
However, in practice, only the components along the propagation direction can be
measured by detecting the change in internal and external angular momentum of atoms.
This indicates that, also in the problem of complete decomposition of
the nucleon spin, only the longitudinal spin decomposition or the helicity
sum rule (not the transverse spin decomposition) would be related to direct
observables.

\section{What is ``Potential Angular Momentum" ?}

We have already pointed out that the difference between the two decompositions
of the nucleon spin is characterized by the two physically inequivalent
orbital angular momenta, i.e. the generalized ``canonical'' OAM and
the ``mechanical'' OAM.
Because the difference of these two OAMs is characterized by the
``potential angular momentum" term \cite{WakaSD10}, a clear understanding of it
is very important for answering the following question, i.e.
``Which of the above two OAMs can be thought as more {\it physical} from
the observational viewpoint ?'' 

Let us first recall the relation,
\begin{equation}
 \bm{L}_{can} \ = \ \bm{L}_{mech} \ + \ \bm{L}_{pot},
\end{equation}
with
\begin{eqnarray}
 \bm{L}_{can} &=& \int \,\psi^\dagger \,\bm{r} \times 
 \left(\,\frac{1}{i} \,\nabla \, - \,g \,\bm{A}_{pure} \,\right) \,\psi \,\,d^3 r , \\
 \bm{L}_{mech} &=& \int \,\psi^\dagger \,\bm{r} \times 
 \left(\,\frac{1}{i} \,\nabla \, - \,g \,\bm{A} \,\right) \,\psi \,\,d^3 r , \label{Lmech} \\ 
 \bm{L}_{pot} &=& \ \ g \,\int \,\psi^\dagger \,\bm{r} \times
 \bm{A}_{phys} \,\psi \,\,d^3 r ,
\end{eqnarray}
which means that the gauge-invariant canonical OAM is obtained as a sum
of the mechanical OAM and the potential OAM defined in Ref.~\refcite{WakaSD10}.
This is clearly different from
the definition adopted by Hatta and Yoshida \cite{HY12}, which is given by
\begin{equation}
 \bm{L}_{mech} \ = \ \bm{L}_{can} \ + \ \bm{L}^\prime_{pot}, \label{Lpot1}
\end{equation}
with
\begin{equation}
 \bm{L}^\prime_{pot} \ = \ - \,g \,\int \,\psi^\dagger \,\bm{r} \times
 \bm{A}_{phys} \,\psi \,\,d^3 r . \label{Lpot2}
\end{equation}
One might think that it is just a matter of sign convention of $\bm{L}_{pot}$
term. It would certainly be so if one is interested only in the difference
between the two OAMs, i.e. the canonical and the mechanical OAMs.
However, if one is interested in the separate physical contents of the two
OAMs, one will find that these two definitions
have quite different physical interpretations.
In fact, as we shall see in the next section,
proper physical interpretation of the two OAMs has a deep
connection with the problem of practical observability of the OAMs
through deep-inelastic-scattering (DIS) measurements.

The reason of Hatta and Yoshida's definition of $\bm{L}^\prime_{pot}$ can
readily be imagined from the following consideration.
In view of the fact that the potential angular momentum contains
the physical component $\bm{A}_{phys}$ of the gluon field, 
$\bm{L}^\prime_{pot}$ is naturally thought to give a measure of the
genuine quark-gluon interaction.
In fact, they showed that $\bm{L}^\prime_{pot}$ is related to the
twist-3 quark-gluon correlation functions.
The spirit of their definition (\ref{Lpot1}) with (\ref{Lpot2}) would then be the
following.
Formally, the expression (\ref{Lmech}) of the mechanical OAM $\bm{L}_{mech}$
contains the full gluon field $\bm{A}$. Then, the subtraction of $\bm{L}^\prime_{pot}$
from $\bm{L}_{mech}$ would work to eliminate the physical part
$\bm{A}_{phys}$ of the gluon field, thereby leading to the generalized
(gauge-invariant) canonical OAM $\bm{L}_{can}$, in which only the pure-gauge
part $\bm{A}_{pure}$ of the gluon is contained. The resultant $\bm{L}_{can}$
is essentially the standard canonical OAM, since $\bm{A}_{pure}$ is an
unphysical gauge degrees of freedom, which can be eliminated eventually.
It also appears that this canonical OAM is perfectly consistent with
free partonic picture of quark orbital motion, as already
emphasized in the paper by Bashinsky and Jaffe \cite{BJ99}. In this picture,
what contains the genuine quark-gluon interaction is $\bm{L}_{mech}$
not $\bm{L}_{can}$.

However, this viewpoint is not necessarily justified. Another totally
different viewpoint would be the following. 
One takes that $\bm{L}_{mech}$ is a quantity with more physical
significance than $\bm{L}_{can}$.
In fact, what appears in the equation of motion of the charged particle
under the presence of the electromagnetic potential is
the mechanical momentum $\bm{P}_{mech}$ and the mechanical OAM
$\bm{L}_{mech}$ not the canonical momentum
$\bm{P}_{can}$ and the canonical OAM $\bm{L}_{can}$ \cite{BookSakurai95}.
Remember that the mechanical OAM is a quantity which is related to the
the coordinate and the velocity of a particle as
$\frac{m \,\bm{r} \times \bm{u}}{\sqrt{1 - \bm{u}^2}}$ with
$\bm{u} = \dot{\bm{r}}$.
To understand the physical meaning of the mechanical OAM, the relativistic
kinematics of the charged particle (electron) is not essential.
It would rather block up transparent understanding of the
physical meaning of the mechanical OAM as well as the relation
between the mechanical OAM and canonical OAM.  
In the following, we therefore consider simpler interacting system of
photons and charged particles with nonrelativistic motion
\cite{BookJR76,BookBLP82,BookCDG89} following the
discussion in Ref.~\refcite{WakaSD12}. The spin
of the charged particle is also discarded, for simplicity.  

The total energy of such a system is given by
\begin{eqnarray}
 H &=& \sum_i \,\frac{1}{2} \,m_i \,\dot{\bm{r}}_i^2 \ + \ 
 \frac{1}{2} \,\int \,[\,\bm{E}^2 + \bm{B}^2 \,] \,d^3 r .
 \label{QED_Hamiltonian}
\end{eqnarray}
Here the 1st and the 2nd terms of the r.h.s. respectively stand for
the mechanical kinetic energy of the charged particles and the
total energy of the electromagnetic fields.

Introducing the transverse-longitudinal decomposition
$\bm{A} (x) = \bm{A}_\perp (x) + \bm{A}_\parallel (x)$ of the
vector potential, the electric field can also be decomposed
into longitudinal and transverse components as
\begin{eqnarray}
 \bm{E} \ = \ \bm{E}_\perp \ + \ \bm{E}_\parallel,
\end{eqnarray}
with
\begin{eqnarray} 
 \bm{E}_\perp &=& - \,\frac{\partial \bm{A}_\perp}{\partial t}, \ \ \ \ \ 
 \bm{E}_\parallel \ = \ - \,\nabla A^0 \ - \ 
 \frac{\partial \bm{A}_\parallel}{\partial t},
\end{eqnarray}
while the magnetic field is intrinsically transverse
\begin{equation}
 \bm{B} \ = \ \nabla \times \bm{A} \ = \ \nabla \times \bm{A}_\perp
 \ = \ \bm{B}_\perp.
\end{equation}
Correspondingly, the photon part of the total energy can be
decomposed into two pieces, i.e. the longitudinal part and the
transverse part, as
\begin{eqnarray}
 H &=& \sum_{i-1}^N \,\frac{1}{2} \,m_i \,\dot{\bm{r}}_i^2 \ + \ 
 \frac{1}{2} \,\int \,\bm{E}_\parallel^2 \,\,d^3 r \ + \ 
 \frac{1}{2} \,\int \,[\,\bm{E}_\perp^2 + \bm{B}_\perp^2 \,] \,\,d^3 r .
\end{eqnarray}
By using the Gauss law $\nabla \cdot \bm{E}_\parallel = \rho$, 
it can be shown that the longitudinal part is nothing but the
Coulomb energy between the charged particles (aside from
the self-energies), so that we can write as
\begin{equation}
 H \ = \  
 \sum_{i = 1}^N \,\frac{1}{2} \,m_i \,\dot{\bm{r}}_i^2 \ \ + \ \ 
 V_{coul} \ \ + \ \ H_{trans} ,
\end{equation}
with
\begin{eqnarray}
 V_{Coul} &=& \frac{1}{4 \,\pi} \,\sum_{i,j = 1 \,(i \neq j)}^N \,
 \frac{q_i \,q_j}{| \bm{r}_i - \bm{r}_j |} , \\
 H_{trans} &=& \frac{1}{2} \,\int \,[\,
 \bm{E}_\perp^2 + \bm{B}_\perp^2 \,] \,\,d^3 r .
\end{eqnarray}

Next, let us consider a similar decomposition of the total momentum.
The total momentum of the system is a sum of the mechanical momentum
of charged particles and the momentum of the photon field as
\begin{eqnarray}
 \bm{P} \ = \ \bm{P}_{mech} \ + \ \bm{P}^\gamma
\end{eqnarray}
with
\begin{eqnarray}
 \bm{P}_{mech} &=& \sum_i \,m_i \,\dot{\bm{r}}_i, \ \ \ \ \ \ 
 \bm{P}^\gamma \ = \ \int \,\bm{E} \times \bm{B} \,\,d^3 r.
\end{eqnarray}

The total momentum of the electromagnetic fields can be decomposed
into longitudinal and transverse parts as
\begin{equation}
 \bm{P}^\gamma \ = \ \bm{P}^\gamma_{long} \ + \ 
 \bm{P}^\gamma_{trans} ,
\end{equation}
with
\begin{eqnarray}
 \bm{P}^\gamma_{long} &=& \int \,\bm{E}_\parallel \times \bm{B}_\perp \,\,d^3 r, \\
 \bm{P}^\gamma_{trans} &=& \int \,\bm{E}_\perp \times \bm{B}_\perp \,\,d^3 r,
\end{eqnarray}
Again, by using the Gauss law, it can be shown that the longitudinal part
$\bm{P}_{long}$ is also expressed as
\begin{equation}
 \bm{P}^\gamma_{long} \ = \ \sum_i \,q_i \,\bm{A}_\perp (\bm{r}_i) \ \equiv \ 
 \bm{P}_{pot} .
\end{equation}
As pointed out in Ref.~\refcite{WakaSD10}, the quantity
$q_i \,\bm{A}_\perp (\bm{r}_i)$ appearing
here is nothing but the {\it potential momentum}
according to the terminology of Konopinski \cite{Konopinski78}.
In the present context, it represents the momentum that associates with the
longitudinal (electric) field generated by the particle $i$.
After these steps, the total momentum of the system is now given as a sum of
three terms as
\begin{equation}
 \bm{P} \ = \ \bm{P}_{mech} \ + \ \bm{P}_{pot} \ + \ \bm{P}^\gamma_{trans}.
\end{equation} 
A delicate question here is the following. 
Which of particles or photons should the potential momentum be
attributed to ? In view of the fact that the potential momentum
term is also thought of as representing the interactions of charged particles
and photons, we realize that it is of the same sort of question as
which of charged particles or photons should the Coulomb energy be
attributed to. To attribute it to charged particle is closer to the concept
of ``action at a distance theory", while to attribute it to electromagnetic
field is closer to the concept of ``action through medium".
If there is no difference between their physical predictions, the choice
is certainly a matter of convenience.
Let us see what happens if we combine the potential momentum term with
the mechanical momentum of charged particles.
To this end, we recall that, under the presence of electromagnetic
potential, the canonical momentum $\bm{p}_i$ of the
charged particle $i$ is given by the equation
\begin{equation}
 \bm{p}_i \ \equiv \ \frac{\partial L}{\partial \dot{\bm{r}}_i}
 \ = \ m_i \,\dot{\bm{r}}_i \ + \ q_i \,\bm{A} (\bm{r}_i) ,
\end{equation}
where $L$ is the Lagrangian corresponding to the Hamiltonian (\ref{QED_Hamiltonian}). 
Using it, the total momentum $\bm{P}$ can be expressed in the following form :
\begin{equation}
 \bm{P} \ = \ \bm{P}_{can} \ + \ \bm{P}^\gamma_{trans}, \label{Pdecompc}
\end{equation}
with
\begin{equation}
 \bm{P}_{can} \ = \ \sum_i \,\left(\bm{p}_i \ - \ q_i \,
 \bm{A}_\parallel (\bm{r}_i) \right) .
\end{equation}
Here use has been made of the relation $\bm{A} (\bm{r}_i) - \bm{A}_\perp (\bm{r}_i)
= \bm{A}_\parallel (\bm{r}_i)$.  
Note that the above $\bm{P}_{can}$ is the generalized (gauge-invariant)
canonical momentum of the charged particle system.

We can carry out a similar manipulation also for the angular momentum.
The total angular momentum of the system is a sum of the mechanical
angular momentum of charged particles and the angular momentum of photon fields as  
\begin{equation}
 \bm{J} \ = \ \bm{L}_{mech} \ + \ \bm{J}^\gamma ,
\end{equation}
with
\begin{eqnarray}
 \bm{L}_{mech} &=& \sum_i \, m_i \,\bm{r}_i \times \dot{\bm{r}}_i, \\
 \bm{J}^\gamma &=& \int \,d^3 r \,\,\bm{r} \times (\bm{E} \times \bm{B}).
\end{eqnarray}
Similarly as before, the total angular momentum of the electromagnetic fields
can be decomposed into longitudinal and transverse parts as
\begin{equation}
 \bm{J}^\gamma \ = \ 
 \bm{J}^\gamma_{long} \ + \ \bm{J}^\gamma_{trans},
\end{equation}
with
\begin{eqnarray}
 \bm{J}^\gamma_{long} &=& \int \,\bm{r} \times 
 \left( \bm{E}_\parallel \times \bm{B}_\perp \right) \,d^3 r, \\
 \bm{J}^\gamma_{trans} &=& \int \,\bm{r} \times
 \left( \bm{E}_\perp \times \bm{B}_\perp \right) \,d^3 r .
\end{eqnarray}
Again, by using the Gauss law, $\bm{J}_{long}$ can also be expressed as
\begin{equation}
 \bm{J}^\gamma_{long} \ = \ \sum_i \,q_i \,\,\bm{r}_i \times \bm{A}_\perp (\bm{r}_i)
 \ \equiv \ \bm{L}_{pot}.
\end{equation}
This is nothing but the potential angular momentum in the terminology
of Ref.~\refcite{WakaSD10}.
Thus, we are led to the following decomposition of the total angular momentum :
\begin{equation}
 \bm{J} \ = \ \bm{L}_{mech} \ + \ \bm{L}_{pot} \ + \ 
 \bm{J}^\gamma_{trans}.
\end{equation}

Again, we have freedom to combine the potential angular momentum with the
mechanical angular momentum of the charged
particle $i$. Noting again the relation $\bm{A} (\bm{r}_i) - \bm{A}_\perp (\bm{r}_i)
= \bm{A}_\parallel (\bm{r}_i)$, we obtain
\begin{equation}
 \bm{J} \ = \ \bm{L}_{can} \ + \ \bm{J}^\gamma_{trans} , \label{Jdecompc}
\end{equation}
with
\begin{equation}
 \bm{L}_{can} \ = \ \sum_i \,\bm{r}_i \times \left(\,\bm{p}_i \ - \ 
 q_i \,\bm{A}_\parallel (\bm{r}_i) \,\right) .
\end{equation}
As expected, this $\bm{L}_{can}$ is just the generalized (gauge-invariant)
canonical OAM of the charged particles.\footnote{This precisely corresponds
to the gauge-invariant canonical OAM appearing in the Chen decomposition.
As is obvious from the above derivation, the gauge
degrees of freedom carried by $\bm{A}_\parallel$ is not introduced
by the artificial prescription of gauge-invariant extension. Rather, it
is a freedom already existing in the original gauge theory,
i.e. in QED.} 
At first sight, simpler-looking appearance of the decomposition
(\ref{Pdecompc}) of the total momentum and the decomposition
(\ref{Jdecompc}) of the total angular mometum
appears to indicate physical superiority of canonical
momentum and the canonical angular momentum over the mechanical ones.
In fact, since $\bm{A}_\parallel (\bm{r}_i)$ is the pure-gauge part of
the photon, which can eventually be eliminated, they are essentially
the momentum and the angular momentum of a free particle.
However, one should recognize the fact that, under the circumstance
where strong electromagnetic potential exists, there cannot be any
free charged particle.
(This observation becomes of more practical importance in the strong-coupled gauge theory like QCD.)
As is clear from the expressions,
\begin{eqnarray}
 \bm{P}_{mech} &=& \sum_i \,m_i \,\dot{\bm{r}}_i \ = \ 
 \sum_i \,m_i \,\bm{v}_i, \\
 \bm{L}_{mech} &=& \sum_i \,m_i \,\bm{r}_i \times \dot{\bm{r}}_i \ = \ 
 \sum_i \,m_i \,\bm{r}_i \times \bm{v}_i ,
\end{eqnarray}
what have natural interpretation as
translational and orbital motions of particles under the presence of
the gauge potential are
the mechanical momentum $\bm{P}_{mech}$ and the mechanical OAM
$\bm{L}_{mech}$ not the canonical momentum $\bm{P}_{can}$ and the
canonical OAM $\bm{L}_{can}$. \cite{BookSakurai95} 
It may sound paradoxical, but in conjunction with
the relation $\bm{P}_{can} = \bm{P}_{mech} \ + \ \bm{P}_{pot}$, and
$\bm{L}_{can} = \bm{L}_{mech} \ + \ \bm{L}_{pot}$, one must say that
what contains extra interaction terms, i.e. the potential momentum and
potential angular momentum, are rather the canonical momentum and
canonical OAM not the mechanical momentum and the mechanical OAM.

One might still suspect that the argument above is just a matter of
philosophy. Naturally, what discriminates physics from philosophy is
the experimental observations. In Sect.7, we will show
that the above-mentioned difference between the canonical OAM and
the mechanical OAM has a crucial influence on their observability
by means of high-energy deep-inelastic-scattering measurements.

\section{On the Relation with Deep-Inelastic-Scattering Observables}

Historically, it was a common belief that the canonical OAMs appearing in
the Jaffe-Manohar decomposition would not correspond to observables,
because they are not gauge-invariant quantities.
This nebulous impression did not change even after a gauge-invariant
version of the Jaffe-Manohar decomposition due to Bashinsky and Jaffe
or of Chen et al. appeared.
However, the situation has changed drastically after Lorc\'{e} and Pasquini
showed that the canonical quark OAM can be related to a certain moment
of a quark distribution function in a phase space, called the Wigner
distribution \cite{LP11}. (A complete classification of the Wigner
distributions for a spin $1/2$ target is given in the paper by
Meissner, Metz, and Schlegel \cite{MMS09}. See also the paper by Lorc\'{e}
and Pasquini for an extension to gluon distributions.\cite{LP13})
Since the longitudinal component of the OAM
arises from the motion of partons in the transverse plane
perpendicular to the nucleon momentum, it is intuitively natural to consider
such generalized distribution functions of partons beyond the collinear
distributions. The relevant quantity here is a phase-space quark distribution
in a longitudinally polarized nucleon :
\begin{eqnarray}
 &\,& \rho^q (x, \bm{k}_\perp, \bm{b}_\perp \,;\,{\cal W})
 \ = \ \int \,\frac{d^2 \Delta_\perp}{(2 \,\pi)^2} \,\,
 e^{\,- \,i \,\bm{\Delta}_\perp \cdot \bm{b}_\perp} \,\,\frac{1}{2} \,
 \int \,\frac{d z^- \,d^2 z_\perp}{(2 \,\pi)^3} \,
 e^{\,i \,(x \,\bar{P}^+ \,z^- - \bm{k}_\perp \cdot \bm{z}_\perp)}
 \nonumber \\
 &\,& \hspace{24mm} \times \  
 \langle P^{\prime +}, \frac{\bm{\Delta}_\perp}{2}, \,S \,|\,
 \bar{\psi} \left( - \,\frac{z}{2} \right) \,\gamma^+ \,
 {\cal W} \,\psi \left( \frac{z}{2} \right) \,|\, P^+,
 - \,\frac{\bm{\Delta}_\perp}{2}, S \rangle \,|_{z^+ = 0} , \hspace{12mm} 
\end{eqnarray}
given as a function of the ordinary longitudinal momentum fraction $x$
($x = k^+ \,/\,\bar{P}^+$ with $\bar{P} = (P^\prime + P) \,/\,2$),
the transverse momentum $\bm{k}_\perp$, the impact parameter $\bm{b}_\perp$.
Here, ${\cal W}$ is a gauge-link, also called the Wilson line, connecting the
two space-time points $z/2$ and $- \,z/2$.
According to them, the Wigner distribution gives a natural definition of
the quark OAM density in the phase-space as follows : 
\begin{equation}
 L_z^q (x, \bm{k}_\perp, \bm{b}_\perp \,;\,{\cal W}) \ = \ 
 (\bm{b}_\perp \times \bm{k}_\perp)_z \,\,
 \rho^q (x,\bm{k}_\perp, \bm{b}_\perp \,;\, {\cal W}) .
\end{equation}
After integrating over $x, \bm{k}_\perp$, and $\bm{b}_\perp$, they arrive at
a remarkable relation, which connects a Wigner distribution with the
quark OAM : 
\begin{eqnarray}
 \langle L^q_z \rangle^{\cal W} &=& \int \,dx \,d^2 k_\perp \,d^2 b_\perp \,
 L^q_z (x, \bm{k}_\perp,\bm{b}_\perp \,;\,{\cal W}) \nonumber \\
 &=& - \,\int \,dx \,d^2 k_\perp \,\,\frac{\bm{k}_\perp^2}{M^2} \,\,
 F^q_{1,4} (x, 0, \bm{k}_\perp^2, 0, 0, {\cal W}),
\end{eqnarray}
where the function $F^q_{1,4}$ is contained in the following structure of
the Wigner distribution $\rho^q$ :
\begin{eqnarray}
 \rho^q (x, \bm{k}_\perp,\bm{b}_\perp \,;\,{\cal W})
 &=& 
 F^q_{1,1} (x,\bm{k}_\perp^2, \bm{k}_\perp \cdot \bm{b}_\perp, 
 \bm{b}_\perp^2 \,;\,{\cal W}) \nonumber \\
 &-& \frac{1}{M^2} \,(\bm{k}_\perp \times \nabla_{\bm{b}_\perp})_z \,
 F^q_{1,4} (x,\bm{k}_\perp^2, \bm{k}_\perp \cdot \bm{b}_\perp, 
 \bm{b}_\perp^2 \,;\,{\cal W}) .
\end{eqnarray}
A delicacy here is that the Wigner distribution $\rho^q$ generally turns out
to depend on the chosen path of the gauge-link ${\cal W}$ connecting the points
$z/2$ and $- \,z/2$.
As shown by a careful study by Hatta \cite{Hatta11} with the choice of a
staple-like gauge-link in the light-front direction, corresponding
to the kinematics of
the semi-inclusive reactions or the Drell-Yan processes, the above quark OAM
turns out to coincide with the (gauge-invariant) canonical quark OAM not
the dynamical OAM :
\begin{equation}
 L^q_{can} \ = \ \langle L^q_z \rangle^{{\cal W} = {\cal W}^{LC}} .
\end{equation}
This observation holds out a hope that the
canonical quark OAM in the nucleon would also be a measurable quantity,
at least in principle.

In a recent paper \cite{CGHLR13}, however, Courtoy et al.  throws a
serious doubt on the practical observability of the Wigner function
$F^q_{1,4}$ appearing in the above intriguing sum rule.
According to them, even though $F^q_{1,4}$ may be nonzero in particular
models and also in real QCD, its observability would be inconsistent with the
following observations :
\begin{itemlist}
 \item it drops out in both the formulation of GPDs and TMDs ;
 \item it is parity-odd, at variance with parity-even structure of more
 familiar TMD Sivers function \cite{Sivers90,Sivers91} ;
 \item it is nonzero only for imaginary values of the quark-proton
 helicity amplitudes.
\end{itemlist}
These observations indicate that $F^q_{1,4}$ would not appear in the cross
section formulas of any DIS processes at least at the leading order approximation.
Anyhow, what is indicated by their arguments is the fact that the
existence of a simple partonic picture of the canonical quark OAM in the
Fock space and its observability are different things. 
It appears to us that this takes a discussion on the
observability of the canonical OAM back to its starting point. 

What about observability of another OAMs, i.e.
the mechanical OAMs, then ?
As pointed out in Sect.2, we already know the relations connecting the
mechanical quark and gluon OAMs to DIS observables : 
\begin{eqnarray}
 L^q_{mech} &=& \frac{1}{2} \,\int \,x \,
 \left[\,H^q (x,0,0) \ + \ E^q (x,0,0) \,\right] \,dx \ - \ 
 \frac{1}{2} \,\int \,\tilde{H}^q (x,0,0) \,dx , \\
 L^G_{mech} &=& \frac{1}{2} \,\int \,x \,
 \left[\,H^G (x,0,0) \ + \ E^G (x,0,0) \,\right] \,dx \ - \ 
 \int \,\tilde{H}^G (x,0,0) \,dx . \ \ \ \ 
\end{eqnarray}
They are indirect relations, however, in the sense that both the quark and
gluon mechanical OAM are obtained only as differences of total angular momenta
and the intrinsic spin parts. It would be nicer, if there is any
sum rule which directly relates the mechanical OAMs to observables.
Fortunately, at least for the quark part, such a relation exists,
as first noticed by Penttinen, Polyakov, and Shuvaev, and Strikman \cite{PPSS00} and later refined by Kiptily and Polyakov \cite{KP04}.
(The same relation has recently been rediscovered by Hatta and Yoshida in their
twist-3 analysis of the nucleon spin contents \cite{HY12}.)
They showed that the mechanical quark OAM $L^q_{mech}$ can be related to
a moment of the twist-3 GPD named $G_2$ as
\begin{equation}
 L^q_{mech} \ = - \,\int \,x \,G^q_2 (x,0,0) \,dx .
\end{equation}
This twist-3 GPD $G^q_2$ appears in the following parametrization of the GPD :
\begin{eqnarray}
 &\,& \frac{1}{2} \,\int \,\frac{d z^-}{2 \,\pi} \,
 e^{\,i \,x \,\bar{P}^+ \,z^-} \,
 \langle P^\prime, S^\prime \,|\, \bar{\psi} \left( - \,\frac{z^-}{2} \right) \,
 \gamma^j \,{\cal W} \,\psi \left( \frac{z^-}{2} \right) \,|\, P, S \rangle
 \nonumber \\
 &\,& = \ \frac{1}{2 \,\bar{P}^+} \,\bar{u} (P^\prime, S^\prime) \,
 \left[\,\frac{\Delta^j_\perp}{2 \,M} \,\,G^q_1 \ + \ \gamma^j \,\,
 (\,H^q + E^q + G^q_2 \,) \right. \nonumber \\
 &\,& \hspace{28mm} \left. + \ \frac{\Delta^j_\perp \,\gamma^+}{P^+} \,\,G^q_3
 \ + \ \frac{i \,\epsilon^{jk}_T \,\Delta^k_\perp \,\gamma^+\,\gamma_5}{P^+} \,
 \,G^q_4 \,\right] \,u (P,S) , 
\end{eqnarray}
where $\bar{P} = (P^\prime + P)/2$, the nucleon is moving in the $z$-direction,
$\bm{\Delta}_\perp$ is the transverse part of $\Delta^\mu$, and the indices
$j,k = 1,2$ are transverse indices. (We recall that the GPD $G_2$ has
some relation to the quantities $\tilde{E}_{2T}$ and $E^3_+$ respectively
discussed in Ref.~\refcite{MMS09} and in Ref.~\refcite{BMK02}.)
Since the mechanical OAM can be given as a $x$-integral of the quantity
$- \,x \,G^q_2 (x,0,0)$, one might be tempted to interpret the latter as a
mechanical OAM density in the Feynman $x$-space. An interesting observation
by Kiptily and Polyakov is the following.
According to them, $G^q_2 (x,0,0)$ consists of the Wandzura-Wilczek
(WW) part and the genuine twist-3 part as follows : 
\begin{equation}
 G^q_2 (x,0,0) \ = \ G^{q,WW}_2 (x,0,0) \ + \ \bar{G}^q_{2} (x,0,0).
\end{equation}
Here, the WW part is represented by the forward limits of the three twist-2 GPDs as
\begin{eqnarray}
 &\,& G^{q,WW}_2 (x,0,0) \ = \ - \,(H^q (x,0,0) + E^q (x,0,0)) \ + \ 
 \frac{1}{x} \,\tilde{H}^q (x,0,0) \nonumber \\
 &\,& \hspace{12mm} + \ \int_x^{\epsilon(x)} \,\frac{dy}{y} \,
 (H^q (y,0,0,) + E^q (y,0,0)) \ - \ \int_x^{\epsilon(x)} \,
 \frac{dy}{y^2} \,\tilde{H}^q (y,0,0) , \hspace{10mm}
\end{eqnarray}
with $\epsilon(x) = x \,/\,|x|$. On the other hand, the 2nd moment of the
genuine twist-3 part of $G^q_2$ is shown
to vanish identically,
\begin{equation}
 \int_{-1}^1 \,x \,\bar{G}^q_2 (x,0,0) \,dx \ = \ 0 . \label{G2_twist3}
\end{equation}
This means that the genuine twist-3 part of $G^q_2$ does not contribute
at all to the net (or integrated) mechanical quark OAM $L^q_{mech}$.
Putting it in another way, the net mechanical quark OAM is determined solely
by three twist-2 GPDs $H^q (x,0,0)$, $E^q (x,0,0)$, and $\tilde{H}^q (x,0,0)$.

Now we recall the discussion in Sect.6 on the relation between
the mechanical and canonical OAMs.
According to the definition of Hatta and Yoshida \cite{HY12},
the mechanical quark OAM
is given as a sum of the canonical quark OAM and the potential angular
momentum as
\begin{equation}
 L^q_{mech} \ = \ L^q_{can} \ + \ L^\prime_{pot}. \label{Lmcp}
\end{equation}
As pointed out there, they showed that the potential angular momentum
$L^\prime_{pot}$ is related to the (`F-type') twist-3 quark-gluon
correlator $\Phi_F (x_1,x_2)$ as
\begin{equation}
 L^\prime_{pot} \ = \ \int \,d x_1 \,d x_2 \,\,{\cal P} \,
 \frac{1}{x_1 - x_2} \, \Phi_F (x_1,x_2) ,
\end{equation}
with ${\cal P}$ denoting a principle value, which means that $L^\prime_{pot}$
is a genuine twist-3 quantity.  
On the other hand, we have seen above that the mechanical quark OAM
$L^q_{mech}$ appearing in the left-hand-side of Eq.(\ref{Lmcp}) is
given by the twist-2
GPDs alone. This dictates that the genuine twist-3 contribution in $L^q_{can}$
and $L^\prime_{pot}$ must cancel each other in their sum.
Is this cancellation accidental ?
Very curiously, if one takes a different viewpoint as advocated in Sect.6,
one can explain the above observation in more natural way.
Namely, by translating the QED argument explained in Sect.6 into the QCD
problem, one observes that the canonical OAM rather emerges as a sum of the
mechanical OAM and the potential angular momentum as
\begin{equation}
 L^q_{can} \ = \ L^q_{mech} \ + \ L_{pot} , \label{Lcmp}
\end{equation}
with the relation $L_{pot} = - \,L^\prime_{pot}$. Now it is no surprise
that the canonical OAM contains the genuine twist-3 part, since it is
given as a sum of the mechanical OAM (given by the twist-2 GPDs alone), and
the genuine twist-3 potential angular momentum. We emphasize that this
interpretation is in perfect harmony with the statement in sect.6, which
tells that what contains the potential angular momentum is the canonical OAM
rather than the mechanical OAM. The consideration above, especially the
relation (\ref{G2_twist3}), also explains naturally
why the mechanical OAM rather than the canonical OAM can be considered as the
{\it physical} one, even though it does not appear to fit with the widespread belief
that the canonical OAM is more compatible with the partonic
interpretation of the OAM angular momentum.  

\begin{figure}[h]
\begin{center}
\includegraphics[width=14.0cm]{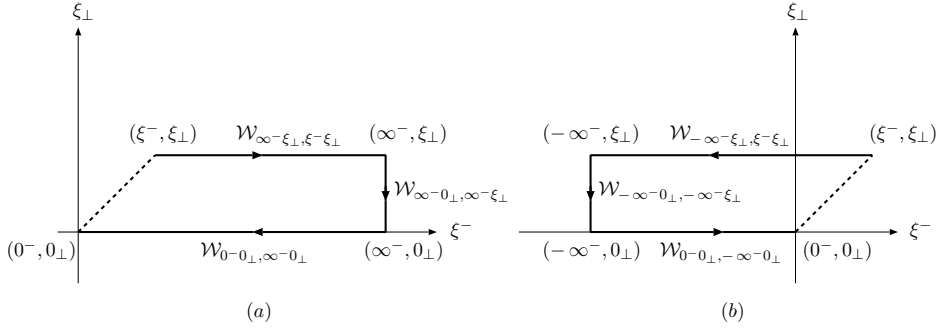}
\caption{\label{Wilson_lines}The future-pointing staple-like Wilson line (left panel)
and the past-pointing staple-like Wilson line, corresponding to the
kinematics of semi-inclusive hadron productions and the Drell-Yan
processes, respectively.}
\end{center}
\end{figure}

The relation between the two kinds of OAMs was also analysed by Burkardt
from a different viewpoint \cite{Burkardt13}.
His attention is paid to a physical interpretation
of the difference between the two OAMs. Also very interesting is his
parallel consideration on the difference between several average transverse
momenta of quarks inside the nucleon. His analysis begins with
the following definitions of the average transverse momenta of
quarks and the longitudinal component of the orbital
angular momenta in terms of Wigner distributions : 
\begin{eqnarray}
 \langle \bm{k}^q_\perp \,\rangle^{\cal W} &=& 
 \int \,dx \,d^2 \bm{b}_\perp \,d^2 \bm{k}_\perp \,\,\bm{k}_\perp \,\,
 \rho^q (x, \bm{b}_\perp, \bm{k}_\perp \,;\, {\cal W}) , \\
 \langle L^q_z \rangle^{\cal W} &=& \int \,dx \,d^2 b_\perp \,d^2 k_\perp \,
 (\bm{b}_\perp \times \bm{k}_\perp)_z \,
 \rho^q (x, \bm{b}_\perp, \bm{k}_\perp \,; \,{\cal W}). 
\end{eqnarray}
These quantities are both dependent on the path of the gauge-kink, since
the Wigner distribution defined by
\begin{eqnarray}
 &\,& \rho^q (x, \bm{k}_\perp, \bm{b}_\perp \,;\,{\cal W})
 \ = \ \int \,\frac{d^2 \Delta_\perp}{(2 \,\pi)^2} \,
 e^{\,- \,i \,\bm{\Delta}_\perp \cdot \bm{b}_\perp} \,\frac{1}{2} \,
 \int \,\frac{d \xi^- \,d^2 \xi_\perp}{(2 \,\pi)^3} \,
 e^{\,i \,(x \,\bar{P}^+ \,\xi^- - \bm{k}\perp \cdot \bm{\xi}_\perp}
 \nonumber \\
 &\,& \hspace{24mm} \times \  
 \langle P^{\prime +}, \frac{\bm{\Delta}_\perp}{2}, \,S \,|\,
 \bar{\psi} \left( 0 \right) \,\gamma^+ \,
 {\cal W} \,\psi \left( \xi \right) \,|\, P^+,
 - \,\frac{\bm{\Delta}_\perp}{2}, S \rangle \,|_{\xi = 0} , \hspace{10mm} 
\end{eqnarray}
is generally dependent on the path connecting the two space-time points
$\xi$ and $0$. Physically interesting paths are the following three.
The first is the future-pointing light-like staple path 
${\cal W}^{+LC}_{0 \xi}$ (see Fig.\ref{Wilson_lines}(a)) corresponding to the
kinematics of semi-inclusive hadron productions. As seen from
Fig.\ref{Wilson_lines}(a), by using the straightline path
${\cal W}^{(sl)}_{\xi^- \bm{\xi}_\perp, \xi^{\prime -} 
\bm{\xi}^{\prime_\perp}}$ connecting the two space-time points by
a straight line, it is represented as ${\cal W}^{+LC}_{0 \xi} \equiv
{\cal W}^{(sl)}_{0^- \bm{0}_\perp, \infty^- \bm{0}_\perp} \,
{\cal W}^{(sl)}_{\infty^- \bm{0}_\perp, \infty^- \bm{\xi}_\perp} \,
{\cal W}^{(sl)}_{\infty^- \bm{\xi}_\perp, \xi^- \bm{\xi}_\perp}$.
The second is the past-pointing light-like staple path
${\cal W}^{-LC}_{0 \xi}$ corresponding to the kinematics of Drell-Yan
processes. It is represented as
${\cal W}^{-LC}_{0 \xi} \equiv
{\cal W}^{(sl)}_{0^- \bm{0}_\perp, - \,\infty^- \bm{0}_\perp} \,
{\cal W}^{(sl)}_{- \,\infty^- \bm{0}_\perp, - \,\infty^- \bm{\xi}_\perp} \,
{\cal W}^{(sl)}_{- \,\infty^- \bm{\xi}_\perp, \xi^- \bm{\xi}_\perp}$, as
illustrated in Fig.\ref{Wilson_lines}(b).
The last is the straightline path connecting the two
space-time points $(0^-, \bm{0}_\perp)$ and $(\xi^-, \bm{\xi}_\perp)$
directly, which gives ${\cal W}^{straight} \equiv 
{\cal W}^{(sl)}_{0^- \bm{0}_\perp, \xi^- \bm{\xi}_\perp}$.

Burkardt primarily concentrated on the difference between the two average
quark transverse momenta and also the difference between the two OAMs, corresponding
to the two gauge-link
paths, i.e. the future-pointing light-like staple path ${\cal W}^{+LC}$
and the straightline path ${\cal W}^{straight}$. When evaluating the
average transverse momentum $\langle k^i_\perp \rangle_{+LC} \equiv
\int \,dx \,d^2 b_\perp \,d^2 k_\perp \,k^i_\perp \,\rho^q (x,\bm{b}_\perp,
\bm{k}_\perp, {\cal W}^{+LC})$, the factor $k^i_\perp$ can be translated into
a derivative $- \,i \,\frac{\partial}{\partial \xi^i_\perp}$ acting
on the operator $\bar{\psi} (0) \,\gamma^+ \,{\cal W}^{+LC}_{0 \xi} \,\psi (\xi)$,
whose matrix element is subsequently evaluated for $\bm{\xi}_\perp = \bm{0}_\perp$.
With use of the familiar derivative formula on the Wilson line \cite{Boer03},
\begin{eqnarray}
 \left. - \,i \,\frac{\partial}{\partial \xi^i_\perp} \,{\cal W}^{+LC}_{0 \xi}
 \right|_{\xi = 0} &=& - \,{\cal W}_{0^- \bm{0}_\perp, \infty^- \bm{0}_\perp} \,
 A^i_\perp (\infty^-, \bm{0}_\perp) \,
 {\cal W}_{\infty^- \bm{0}_\perp, 0^- \bm{0}_\perp}
 \nonumber \\
 &+& \int_{0^-}^\infty \,d z^- \,{\cal W}_{0^- \bm{0}_\perp, z^- \bm{0}_\perp} \,
 \partial^i \,A^+ (z^-, \bm{0}_\perp) \,
 {\cal W}_{z^- \bm{0}_\perp, 0^- \bm{0}_\perp} , \ \ \ \ \ \ 
\end{eqnarray}
one readily obtains
\begin{eqnarray}
 \langle k^i_\perp \rangle^{+LC} \ &=& \ {\cal N} \,
 \int \,d^3 r \,\,\langle P S \,|\,
 \bar{\psi} (\bm{r}) \,\gamma^+ \,\left(\, \frac{1}{i} \,\,\nabla^i_\perp - 
 g \,A^i_\perp (r^-,\bm{r}_\perp) \right. \nonumber \\
 &\,& - \,\left. \int_{r^-}^\infty \,d z^- \,
 {\cal W}_{r^- \bm{r}_\perp, z^- \bm{r}_\perp} \,
 F^{+ i} (z^-, \bm{r}_\perp) \,{\cal W}_{z^- \bm{r}_\perp, r^- \bm{r}_\perp} \,
 \right) \psi (\bm{r}) \,|\, P S \rangle , \ \ \ \ \ \ \ \ 
\end{eqnarray}
with ${\cal N} = 1 \,/\, \langle P S \,|\, P S \rangle$.
Similarly, it can be shown that the average transverse momentum corresponding to
the straightline path ${\cal W}^{straight}$ is given by
\begin{eqnarray}
 &\,& \langle k^i_\perp \rangle^{straight} \nonumber \\
 &\,& \ = \ {\cal N} \,\int \,d^3 r \,\,
 \langle P S \,|\,\bar{\psi} (\bm{r}) \,\gamma^+ \,\left(\,
 \frac{1}{i} \,\,\nabla^i_\perp - g \,A^i_\perp (r^-, \bm{r}_\perp) \,
 \right) \,\psi (\bm{r}) \,|\, P S \rangle . \ \ \ \ 
\end{eqnarray}
The difference between these two quantities is therefore given by
\begin{eqnarray}
 &\,& \langle k^i_\perp \rangle^{+LC} \ - \ \langle k^i_\perp \rangle^{straight}
 \ = \ - \,\,{\cal N} \,\int \,d^3 r \,\,\langle P S \,|\,
 \bar{\psi} (\bm{r}) \,\gamma^+ \nonumber \\
 &\,& \hspace{15mm} \times \ \int_{r^-}^\infty \,
 {\cal W}_{r^- \bm{r}_\perp, z^- \bm{r}_\perp} \,F^{+ i} (z^-, \bm{r}_\perp) \,
 {\cal W}_{z^- \bm{r}_\perp, r^- \bm{r}_\perp} \,\psi (\bm{r}) \,|\, P S \rangle
 , \ \ \ \ \label{Trans_force}
\end{eqnarray}
although one can show that $\langle k^i_\perp \rangle^{straight}$
vanishes by time-reversal invariance. 

As pointed out by Burkardt, the quantity on the r.h.s. is nothing but the
well-known Qiu-Sterman matrix element \cite{QS91}.
According to him \cite{Burkardt13}, this quantity
can be interpreted as the change of transverse momentum for the struck
quark as it leaves the target after being struck by the virtual photon
in the semi-inclusive DIS processes. The legitimacy of this interpretation
can most easily be seen by taking the light-cone gauge.
In this gauge, the Wilson lines along the light-like direction become unity
and the relevant component of the field-strength tensor,
for example, $F^{+y}$, reduces to
\begin{eqnarray}
 - \,\sqrt{2} \,g \,F^{+ y} &=& - \,g \,F^{0 y} \ - \ g \,F^{z y}
 \ = \ g \,(E^y \ - \ B^x) \nonumber \\
 &=& g \,\left[\bm{E} \ + \ (\bm{v} \times \bm{B})\right]^y,
\end{eqnarray}
which represents the $y$-component of the color Lorentz force acting on a
particle that moves with the velocity of light in the $- \,z$ direction,
i.e. $\bm{v} = (0,0,- \,1)$, that is the direction of the momentum
transfer in the semi-inclusive DIS reactions. This motivates him the
semiclassical interpretation of the matrix element of (\ref{Trans_force}) as the
average transverse momentum of the ejected quark generated by
the average color-Lorentz force from the spectator as it leaves the target.

A similar analysis can be carried out also for the quark OAMs, although
it needs an extra care. That is, when one evaluates
$\langle L^q_z \rangle^{+LC} = \int \,dx \,d^2 b_\perp \,d^2 k_\perp \,
(\bm{b}_\perp \times \bm{k}_\perp)^z \,\rho^q (x, \bm{b}_\perp, \bm{k}_\perp
\,; \,{\cal W}^{+LC})$, the factor $\bm{b}_\perp$ can be translated into a derivative $- \,i \,\frac{\partial}{\partial \Delta^i_\perp}$ acting on
the matrix element $\langle P^\prime S^\prime \,|\,\bar{\psi}(0) \,
\gamma^+ \,{\cal W}^{+LC}_{0 \xi} \,\psi(\bm{r}) \,|\, P S \rangle = 
\langle \bar{P} - \frac{\Delta}{2}, S^\prime \,|\,\bar{\psi}(0) \,
\gamma^+ \,{\cal W}^{+LC}_{0 \xi} \,\psi(\bm{r}) \,|\, \bar{P} 
+ \frac{\Delta}{2}, S \rangle$
contained in the definition of the Wigner distribution
$\rho^q (x, \bm{b}_\perp, \bm{k}_\perp \,;\, {\cal W}^{+LC})$.
Hatta carried out this nontrivial operation by making use of a parametrization
of the above nucleon matrix element in terms of the twist-3 quark-gluon
correlation \cite{Hatta12,HY12}. Using the notation of
Burkardt \cite{Burkardt13}, this gives
\begin{eqnarray}
 &\,& \!\!\!\! \langle L^q_z \rangle^{+LC} \nonumber \\
 = &\,& {\cal N} \,\int \,d^3 r \,\langle P S \,|\,\bar{\psi} (\bm{r}) \,
 \gamma^+ \,\left\{ \,\left[\,\bm{r} \times 
 \left(\,\frac{1}{i} \,\nabla - \,g \,\bm{A} \,\right) \,\right]^z \right.
 \ - \ \int_{r^-}^\infty \,d z^- \,
 {\cal W}_{r^- \bm{r}_\perp, z^- \bm{r}_\perp} \,
 \nonumber \\
 &\,& \hspace{8mm} \left. \times \ g 
 \left( x \,F^{+ y} (z^-, \bm{r}_\perp) - y \,F^{+ x} (z^-, \bm{r}_\perp) \right) \,
 {\cal W}_{r^- \bm{r}_\perp, r^- \bm{r}_\perp} \,\right\} \,\psi (\bm{r}) \,
 |\, P S \rangle . \ \ \ \ \ \ \ 
\end{eqnarray}
Similarly, as first noticed by Ji, Kiong, and Yuan \cite{JXY12B},
the OAM corresponding
to the straightline path ${\cal W}_{straight}$ is given by
\begin{eqnarray}
 \langle L^q_z \rangle^{straight} 
 \ = \ {\cal N} \,\int \,d^3 r \,\langle P S \,|\,\bar{\psi} (\bm{r}) \,
 \gamma^+ \left[\,\bm{r} \times 
 \left(\,\frac{1}{i} \,\nabla - \,g \,\bm{A} \,\right) \right]^z
 \psi (\bm{r}) \,|\, P S \rangle .
 \ \ \  
\end{eqnarray}
One therefore finds for the difference
\begin{eqnarray}
 &\,& \langle L^q_z \rangle^{+LC} \ - \ \langle L^q_z \rangle^{straight} 
 \ = \ - \,{\cal N} \,\int \,d^3 r \,
 \langle P S \,|\, \bar{\psi} (\bm{r}) \,\gamma^+ \,
 \int_{r^-}^\infty \,d z^- \,{\cal W}_{r^- \bm{r}_\perp,z^- \bm{r}_\perp}
 \nonumber \\
 &\,& \hspace{15mm} \times \ g \,\left( x \,F^{+ y} (z^-, \bm{r}_\perp) - 
 y \,F^{+x} (z^-, \bm{r}_\perp) \right) \,
 {\cal W}_{z^- \bm{r}_\perp, r^- \bm{r}_\perp} \,\psi (\bm{r}) \,|\, P S \rangle
 . \ \ \ \ \ \ \ \ 
\end{eqnarray}
Analogous to the previous semiclassical interpretation of 
$- \,g \,F^{+i} (r^-,\bm{r}_\perp)$ as the transverse force acting on
the active quark along its trajectory, Burkardt gave an interpretation that
\begin{equation}
 T^z (r^-, \bm{r}_\perp) \ \equiv \ - \,g \,
 \left(\, x \,F^{+y} (r^-,\bm{r}_\perp)
 \ - \ y \,F^{+x} (r^-, \bm{r}_\perp) \,\right),
\end{equation}
represents the $z$-component of the {\it torque} that acts on a particle moving
with the velocity of light in the $- \,z$ direction - the direction in which the
ejected quark moves. Consequently, the difference between the two OAMs, i.e.
$\langle L^q_z \rangle^{+LC}$ and $\langle L^q_z \rangle^{straight}$
is interpreted as the change of the OAM
as the quark moves through the color field created by the
spectators. We point out that these two OAMs $\langle L^q_z \rangle^{+LC}$
and $\langle L^q_z \rangle^{straight}$
are nothing but the ``canonical'' and ``mechanical'' OAMs, 
$L_{can}$ and $L_{mech}$, respectively.

The analysis of Burkardt is limited only to the difference between
the two average transverse momenta and the difference between the
two OAMs. We find it very interesting to reconsider his analysis from
a different viewpoint.
Let us start with the following relations : 
\begin{eqnarray}
 &\,& \langle k^i_\perp \rangle^{\pm LC} \ = \ {\cal N} \,
 \int \,d^3 r \,\langle P S \,|\,
 \bar{\psi} (\bm{r}) \,\gamma^+ \,\left(\, \frac{1}{i} \,\nabla^i_\perp - 
 g \,A^i_\perp (r^-,\bm{r}_\perp) \right. \nonumber \\
 &\,& \hspace{15mm} \left. - \int_{r^-}^{\pm \infty} d z^- \,
 {\cal W}_{r^- \bm{r}_\perp, z^- \bm{r}_\perp} \,
 F^{+ i} (z^-, \bm{r}_\perp) \,{\cal W}_{z^- \bm{r}_\perp, r^- \bm{r}_\perp} \,
 \right) \psi (\bm{r}) \,|\, P S \rangle . \ \ \ \ \ \ \ \ 
\end{eqnarray}
and
\begin{eqnarray}
 &\,& \!\!\!\! \langle L^q_z \rangle^{\pm LC} \nonumber \\
 = &\,& {\cal N} \,\int \,d^3 r \,\langle P S \,|\,\bar{\psi} (\bm{r}) \,
 \gamma^+ \,\left\{ \,\left[\,\bm{r} \times 
 \left(\,\frac{1}{i} \,\nabla - \,g \,\bm{A} \,\right) \,\right]^z \right.
 \ - \ \int_{r^-}^{\pm \infty} \,{\cal W}_{r^- \bm{r}_\perp, z^- \bm{r}_\perp} \,
 \nonumber \\
 &\,& \hspace{6mm} \left. \times \ g \,
 \left( x \,F^{+ y} (z^-, \bm{r}_\perp) - y \,F^{+ x} (z^-, \bm{r}_\perp) \right) \,
 {\cal W}_{z^- \bm{r}_\perp, r^- \bm{r}_\perp} \,\right\} \,\psi (\bm{r}) \,
 |\, P S \rangle . \ \ \ \ \ \ \ 
 \label{LpmLC}
\end{eqnarray}
Here, we have given the average transverse momentum and the OAM
not only for the future-pointing light-like staple path
but also for the past-pointing one.
As pointed out by Hatta, the two OAMs 
$\langle L^q_z \rangle^{\pm LC}$ are actually shown
to coincide due to parity and time-reversal (PT) symmetry,
and they
can be identified with the (gauge-invariant) canonical OAM as
\begin{equation}
 \langle L^q_z \rangle^{+ LC} \ = \ 
 \langle L^q_z \rangle^{- LC} \ = \ \frac{1}{2} \,
 (\langle L^q_z \rangle^{+LC} \ + \ \langle L^q_z \rangle^{-LC})
 \ = \ L_{can} . \label{PT_L}
\end{equation}
In fact, the quantity appearing in the 2nd term of
$\langle L^q_z \rangle^{\pm LC}$ can be written as
\begin{eqnarray}
 &\,& - \,\int_{r^-}^{\pm \infty} \,d z^- \,
 {\cal W}_{r^- \bm{r}_\perp, z^- \bm{r}_\perp}
 \,F^{+ \mu} (z^-,\bm{r}_\perp) \,{\cal W}_{z^- \bm{r}_\perp, r^- \bm{r}_\perp}
 \nonumber \\
 &\,& = \ - \,\int \,d z^- \,\kappa (z^- - r^-),
 {\cal W}_{r^- \bm{r}_\perp, z^- \bm{r}_\perp} \,
 F^{+ \mu} (z^-,\bm{r}_\perp) \,{\cal W}_{z^- \bm{r}_\perp, r^- \bm{r}_\perp} ,
\end{eqnarray}
with use of the functions
\begin{equation}
 \kappa (z^-) \ = \ \pm \,\theta (\pm \,z^-) ,
\end{equation}
depending on the two choices of path. The above quantity precisely coincides with
the physical component of the gluon defined by Hatta \cite{Hatta11},
\begin{eqnarray}
 &\,& A^\mu_{phys} (r^-,\bm{r}_\perp) \nonumber \\
 &\,& = - \,\int \,d z^- \,\kappa (z^- - r^-) \,
 {\cal W}_{r^- \bm{r}_\perp, z^- \bm{r}_\perp} \,
 F^{+ \mu} (z^-,\bm{r}_\perp) \,{\cal W}_{z^- \bm{r}_\perp, r^- \bm{r}_\perp} .
 \ \ \ \ \ 
\end{eqnarray}
Plugging this into (\ref{LpmLC}), one thus obtains
\begin{eqnarray}
 \langle L^q_z \rangle^{\pm LC} &=& {\cal N} \,
 \int \,d^3 r \,\langle P S \,|\,\bar{\psi} (\bm{r}) \,
 \gamma^+ \,\left\{\,\left[ \bm{r} \times 
 \left( \frac{1}{i} \,\nabla - g \,\bm{A} (r^-,\bm{r}_\perp) \right) \right]^z
 \right. \nonumber \\
 &\,& \left. \hspace{24mm} + \ \left( \bm{r} \times 
 \bm{A}_{phys} (r^-,\bm{r}_\perp) \right)^z \,\right\} \,
 \psi (\bm{r}) \,|\, P S \rangle  \nonumber \\
 &=& {\cal N} \,\int d^3 r \,\langle P S \,|\,\bar{\psi} (\bm{r}) \,\gamma^+ \,
 \left[ \bm{r} \times \left( \frac{1}{i} \,\nabla - g \,\bm{A}_{pure}
 (r^-, \bm{r}_\perp) \right) \right]^z \,\psi (\bm{r}) \,|\, P S \rangle
 . \ \ \ \ \ \nonumber \\
 &=& L_{can} . 
\end{eqnarray}
The r.h.s. of this equation in fact reproduces the theoretical expression for
the gauge-invariant canonical momentum $L_{can}$.
What is important here is that, because of the equality (\ref{PT_L}), 
the definition of the canonical OAM
is independent of the two choice of the light-like paths relevant to the
two physical DIS processes.

As we shall see below, however, this is not the case for the average
transverse momenta. In fact, exactly in the same way as the manipulation
above, one can show that
\begin{eqnarray}
 \langle k^i_\perp \rangle^{\pm LC} &=& {\cal N} \,
 \int \,d^3 r \,\langle P S \,|\,\bar{\psi} (\bm{r}) \,
 \left[ \gamma^+ \,\left( \frac{1}{i} \,\nabla^i_\perp - g \,
 A^i (r^-,\bm{r}_\perp) \right) \right. \nonumber \\
 &\,& \hspace{38mm} \left. + \ g \,A^i_{phys} (r^-, \bm{r}_\perp) \,\right] \,
 \psi (\bm{r}) \,|\, P S \rangle . \ \ \ 
\end{eqnarray}
This therefore gives
\begin{eqnarray}
 &\,& \langle k^i_\perp \rangle^{\pm LC} \nonumber \\
 &\,& = \ {\cal N} \,\int d^3 r \,\langle P S \,|\,\bar{\psi} (\bm{r}) \,
 \gamma^+ \left( \frac{1}{i} \,\nabla^i_\perp - g \,
 A^i_{pure} (r^-,\bm{r}_\perp) \right)  
 \psi (\bm{r}) \,|\, P S \rangle . 
 \ \ \ \ \ \ 
\end{eqnarray}
Formally, the r.h.s. of this relation is the defining equation
of canonical transverse momentum $\langle k^i_\perp \rangle_{can}$.
A problem here is that the two average transverse momenta
$\langle k^i_\perp \rangle^{\pm LC}$ do not agree with each other.
In fact, from PT symmetry, they
have opposite sign with equal magnitude \cite{Collins02}
\begin{equation}
 \langle k^i_\perp \rangle^{- LC} \ = \ - \,\langle k^i_\perp \rangle^{+ LC} .
\end{equation}
This means that the definition of the average transverse canonical
momentum is not universal. Putting it in another way, while the
potential angular momentum $L^\prime_{pot}$ defined by
\begin{equation}
 L^\prime_{pot} \ \equiv \ L_{mech} \ - \ L_{can},
\end{equation}
may basically be a universal quantity, the potential momentum defined by
\begin{equation}
 \langle k^i_\perp \rangle_{pot} \ \equiv \ \langle k^i_\perp \rangle_{mech}
 \ - \ \langle k^i_\perp \rangle_{can} 
\end{equation}
is not a path-independent quantity. Since $\langle k^i_\perp \rangle_{mech} = 0$
by time-reversal symmetry, one can also say that the potential
momentum has just opposite
sign with equal magnitude for semi-inclusive and Drell-Yan processes.
One natural possibility of defining the canonical transverse momentum
might be to take an average, i.e.
\begin{equation}
 \langle k^i_\perp \rangle_{can} \ \equiv \ \frac{1}{2} \,
 \left( \langle k^i_\perp \rangle^{+LC} \ + \ 
 \langle k^i_\perp \rangle^{-LC} \right),
\end{equation}
which gives
\begin{equation}
 \langle k^i_\perp \rangle_{can} \ = \ 0.
\end{equation}
It however does not correspond to a single DIS process, thereby making
its physical meaning obscure. This consideration in turn indicates
somewhat peculiar nature of the canonical momentum
corresponding to the transverse motion.

What is curious here is the origin of the difference between the case
of the average transverse momentum and that of the OAM.
A plausible reason might be the following.
In the argument of OAM, we are considering its longitudinal component,
i,e. the component along the direction of nucleon momentum.
On the other hand, in the case of average transverse momentum, we are
dealing with the components perpendicular to the direction of nucleon
momentum. To understand the significance of this difference, we recall here
more familiar discussion on the longitudinal momentum fractions of
quarks and gluons. It starts with the standard gauge-invariant
decomposition of the QCD energy momentum tensor given as follows :   
\begin{equation}
 T^{\mu \nu} \ = \ T^{\mu \nu}_q \ + \ T^{\mu \nu}_G , \label{EMtensor_mech}
\end{equation}
with
\begin{eqnarray}
 T^{\mu \nu}_q &=& \frac{1}{2} \,\bar{\psi} \,\left(\, \gamma^\mu \,i \,D^\nu
 \ + \ \gamma^\nu \,i \,D^\mu \,\right) \,\psi , \\
 T^{\mu \nu}_G &=& 2 \,\mbox{Tr} \,[\,F^{\mu \alpha} \,F_{\alpha}{}^\nu \,]
 \ + \ \frac{1}{2} \,\mbox{Tr} \,F^2 .
\end{eqnarray}
The quark part of the above QCD energy-momentum tensor is the famous
Belinfante symmetric tensor, or the mechanical energy-momentum tensor
of quarks. Eq.(\ref{EMtensor_mech}) therefore gives the mechanical
decomposition of the QCD energy momentum tensor.

On the other hand, the ``seemingly'' covariant version of 
the canonical decomposition, takes the following form : 
\begin{equation}
 T^{\mu \nu} \ = \ T^{\prime \mu \nu}_q \ + \ T^{\prime \mu \nu}_G ,
 \label{EMtensor_can}
\end{equation}
with
\begin{eqnarray}
 T^{\prime \mu \nu}_q &=& \frac{1}{2} \,\bar{\psi} \,
 \left(\, \gamma^\mu \,i \,D^\nu_{pure}
 \ + \ \gamma^\nu \,i \,D^\mu_{pure} \,\right) \,\psi , \\
 T^{\prime \mu \nu}_G &=& -  \,\mbox{Tr} \,
 [\,F^{\mu \alpha} \,D^\nu_{pure} A_{\alpha,phys} \ + \ 
 F^{\nu \alpha} \,D^\mu_{pure} \,A_{\alpha,phys} \,]
 \ + \ \frac{1}{2} \,\mbox{Tr} \,F^2 .
\end{eqnarray}
Note that the quark part of this decomposition stands for the canonical
energy-momentum tensor, aside from unphysical gauge degrees of freedom.
Eq.(\ref{EMtensor_can}) therefore gives the canonical decomposition of
the QCD energy momentum tensor.
What do these two different decompositions predict for the momentum sum
rule of QCD ? Utilizing the freedom of gauge choice, one can take the
light-cone gauge ($A^+ = 0$). In this case, we can set
\begin{eqnarray}
 &\,& A^+_{phys} \ \rightarrow \ 0, \hspace{10mm} A^+_{pure} \ \rightarrow \ 0, \\
 &\,& D^+ \ \equiv \ \partial^+ \,- \,i \,g \,A^+ 
 \ \rightarrow \ \partial^+ , \ \ \  
 D^+_{pure} \ \equiv \ \partial^+ \,- \,i \,g \,A^+_{pure} 
 \ \rightarrow \ \partial^+, 
 \ \ \ \ \\
 &\,& F^{+ \alpha} \ = \ \partial^+ A^\alpha \,- \, \partial^\alpha A^+ \ - \ 
 \,g \,[A^+, A^\alpha] \ \rightarrow \, \partial^+ A^\alpha .
\end{eqnarray}
Consequently, $T^{++}$ component in {\it either} of the above two decompositions
reduce to the following simple form,
\begin{equation}
 T^{++} \ = \ i \,\psi^\dagger_+ \,\partial^+ \,\psi_+ \ + \ 
 \mbox{Tr} \,(\partial^+ \bm{A}_\perp)^2 ,
\end{equation}
where $\psi^+ \equiv \frac{1}{2} \,\gamma^+ \,\gamma^- \,\psi$ with
$\gamma^\pm = (\gamma^0 \pm \gamma^3) \,/\,\sqrt{2}$ is the so-called
good component of the quark field. As emphasized by Jaffe many years
ago \cite{Jaffe01},
interaction-dependent part drops in the light-cone gauge and infinite-momentum
frame. Thus, from
\begin{equation}
 \langle P_\infty \,|\, T^{++} \,|\,P_\infty \rangle \,/\, 
 2 \,(P^+_\infty)^2 \ = \ 1,
\end{equation}
we are led to the standard momentum sum rule of QCD given as
\begin{equation}
 \langle x \rangle^q \ + \ \langle x \rangle^G \ = \ 1 .
\end{equation}
Note that even the canonical decomposition gives this standard sum rule,
contrary to the claim in the original paper by Chen et al. \cite{Chen09}.
The point is that the difference between
the canonical energy momentum tensor
\begin{equation}
 T^{\prime ++}_q \ = \ \frac{1}{2} \,\bar{\psi} \,(\gamma^+ \,i \,\partial^+
 \ + \ \gamma^+ \,i \,\partial^+ \,) \,\psi ,
\end{equation}
and the mechanical energy momentum tensor
\begin{equation}
 T^{++}_q \ = \ \frac{1}{2} \,\bar{\psi} \,(\gamma^+ \,i \,D^+ \ + \ 
 \gamma^+ \,i \,D^+ \,) \,\psi ,
\end{equation}
does not have effect on the longitudinal momentum sum rule after taking
the light-cone gauge ($A^+ = 0$).
We have seen that this simple argument does not hold for the average transverse momentum, and that there is no universal definition of the canonical
momentum in this case.
It seems to us that this once again shed light on ``unphysical''
(or ``mathematical'') nature of the idea of canonical momentum, at least
in its most general context.
By some deep reason\footnote{We conjecture that it is not unrelated
to the fact that only the component of the gauge field along the
propagation direction can be decomposed into the physical and
pure-gauge parts in a gauge- and frame-independent way.}, 
such a discrepancy does not occur for
the longitudinal component of the OAMs. Nevertheless, by drawing on all
the arguments in this sections, we feel that what has closer
relationship with physical observables is the mechanical OAM rather than
the canonical OAM.

After overviewing various aspects of the gauge-invariant nucleon spin
decomposition problem, we think it useful to revisit the consideration
on the Stueckelberg symmetry given in sect.3.
According to Lorc\'{e} \cite{Lorce12,Lorce13A,Lorce13B,Lorce13C,Lorce14},
the Stueckelberg symmetry dictates existence of infinitely
many decomposition of the gluon field into its physical and pure-gauge component,
and this in turn leads to infinitely many GIEs of Jaffe-Manohar decomposition
of the nucleon spin. 
However, summing up the consideration so far, it appears that the
gauge symmetry plays only the secondary role in the existence of
plural forms of nucleon spin decomposition. 
First, we now understand that there are in principle infinitely many
definitions of relativistic spin operator (we are supposing here, 
for example, the existence of many definitions of transverse spin), 
the origin of which can be attributed to
the relativity not the gauge symmetry.\footnote{As already pointed out before,
the ultimate origin of it
can be traced back to the fact that successive operations of Lorentz boosts
generate spin rotation.}
Next, suppose that we are considering one of these
spin operators, and that it does not have manifest gauge-invariance.
Such an operator can readily (or trivially) be made gauge-invariant
by making use of Wilson lines.
This is the point at which the path-dependence arises in the
definition of the spin operators.
Nevertheless, from the context  of physical application, there are
actually restricted numbers of paths, at least at
the dominant order of twist expansion. (We are supposing here, for instance,
the future-pointing and past-pointing light-like staple paths, which appears
in the definitions of TMDs and/or Wigner distributions.)
According to Lorc\'{e}'s viewpoint, they might be called different GIEs.
However, we do not necessarily need to use such a conceptually
strange notion like the GIEs. In fact, plainer interpretation would be that
there are simply two different definitions of relativistic spin observables
(or quasi observables), both of which correspond to
different experimental settings.
Despite these general statements, exceptional features of the longitudinal
nucleon spin sum rule should not be forgotten.
First, the relativity does not interude a unique definition of the sum rule,
essentially because the helicity of the massless gluon is a Lorentz-invariant
concept. Second, there is no essential path-dependence in
the definition of the gluon spin operator, which can be defined as the
1st moment of a collinear distribution function.
(The relevant path here is just a light-like straightline path.) 
This enables us to get gauge- as well as Lorentz-frame-independent
decompositions of the longitudinal nucleon spin in a traditional sense.

\section{Lattice QCD Studies of Nucleon Spin Contents}

As pointed out in Sect.1, among the 4 pieces of the longitudinal
nucleon spin decompositions, only the intrinsic quark spin contribution
has been fairly precisely determined through experiments.
Empirical information on the other parts is still very poor.
Fortunately, there have been a great progress from the theoretical side.
That is, we can now get valuable information from the lattice QCD simulations,
which provides us with a powerful tool for handling nonperturbative QCD.
Over the last few years, the two lattice QCD collaborations carried out
extensive studies on the nucleon spin contents, although within
the so-called quenched approximation \cite{LHPC08,LHPC10,Hagler10,
QCDSF-UKQCD06,QCDSF-UKQCD07}.
The basis of these analyses is the well-known Ji sum rule : 
\begin{eqnarray}
 J^q &=& \frac{1}{2} \,\left[\, A^q_{20} (0) \ + \ B^q_{20} (0) \,\right], \\
 J^G &=& \frac{1}{2} \,\left[\, A^G_{20} (0) \ + \ B^G_{20} (0) \,\right],
\end{eqnarray}
with $A^q_{20} (0)$ and $B^q_{20} (0)$ being the forward limits of the
so-called generalized form factors, which are related to the 
forward limits of the unpolarized GPDs as
\begin{eqnarray}
 A^q_{20} (0) &=& \int_{-1}^1 \,x \,H^q (x,0,0) \,dx , \\
 B^q_{20} (0) &=& \int_{-1}^1 \,x \,E^q (x,0,0) \,dx ,
\end{eqnarray}
and similarly for the gluon part. Here, we confine to the quark part,
since the study of the gluon part is still a difficult challenge
even for the lattice QCD. 
The lattice QCD simulations concentrates on evaluating the four
quantities $A^{u \pm d}_{20} (0)$, $B^{u \pm d}_{20} (0)$, which is
necessary to get separate knowledge on $J^u$ and $J^d$.
Once the total angular momentum $J^q$ of the quark with a particular
flavor is known, the quark OAM is obtained through the relation,
\begin{equation}
 L^q \ = \ J^q \ - \ \frac{1}{2} \,\Delta \Sigma^q .
\end{equation}
Here, $\Delta \Sigma^q$ is the 1st moment of the familiar longitudinally
polarized quark distribution,
\begin{equation}
 \Delta \Sigma^q \ = \ \int_{-1}^1 \,\Delta q (x) \,dx,
\end{equation}
with the corresponding flavor $q$. Needless to say, the quark
OAM obtained in this way corresponds to the ``mechanical'' OAM not the
``canonical'' OAM.

Shown in Table \ref{LHPC} are the lattice QCD predictions for $2 \,J \,\equiv \,
2 \,(J^u + J^d + J^s)$, $\Delta \Sigma \,\equiv \,\Delta u + \Delta d + \Delta s$,
and $2 \,L \,\equiv \,2 \,(L^u + L^d + L^s)$ by the
LHPC \cite{LHPC08,LHPC10,Hagler10} and
QCDSF-UKQCD groups \cite{QCDSF-UKQCD06,QCDSF-UKQCD07}.
For the sake of comparison,
we also show here the corresponding predictions of the chiral quark soliton
model (CQSM) evolved to the energy scale of
$Q^2 = 4 \,\mbox{GeV}^2$  \cite{WN06,WN08}, which corresponds
to the renormalization scale of lattice QCD calculations.
The prediction of the CQSM is shown,
because it is a particularly
successful model of the nucleon structure
functions \cite{WGR96,WGR97,GRW98,DPPPW96,DPPPW97,PPPGWW99,WK98,WK99,
Waka03A,Waka03B}.
In particular, it is almost only one effective model of the nucleon.
which is able to explain the observed smallness of the quark spin fraction
of the nucleon without any fine-tuning.
Moreover, this unique prediction of the model is inseparably
connected with its basic physical picture of the nucleon as a rotating
hedgehog, which in turn predicts fairly large orbital angular momentum of
quarks  \cite{WY91,WW00}. 
As seen from the table, this interesting prediction of the CQSM
does not seem to be supported by the lattice QCD predictions of
the LHPC and the QCDSF-UKQCD Collaborations.
The results of both groups show that
both the $u$- and $d$-quark OAMs carry sizable amount
(nearly $20 \,\%$) of the nucleon spin. However, their contributions
to the net nucleon spin tend to cancel in such a way that
\begin{equation}
 2 \,L^u \ \simeq \ - \,0.2, \ \ \ 2 \,L^d \ \simeq \ + \,0.2, \ \ \ 
 2 \,L^{u+d} \ \simeq \ 0.
\end{equation}

\renewcommand{\arraystretch}{1.0}
\begin{table}[hp]
\tbl{Lattice QCD predictions for the nucleon spin contents by the LHPC
\cite{LHPC08,LHPC10,Hagler10} and QCDSF-UKQCD Collaborations 
\cite{QCDSF-UKQCD06,QCDSF-UKQCD07}.
Also shown for comparison are the predictions of the chiral quark
soliton model evolved to the renormalization scale of lattice QCD.}
{\begin{tabular}{@{}cccc@{}} \toprule
\hspace{16mm} & \ \ LHPC \ \ & \ \ QCDSF-UKQCD \ \ & \ 
CQSM ($Q^2 = 4 \,\mbox{GeV}^2$) \ \\ \colrule
$2 \,J$ & 0.426(48) & 0.452(26) & 0.676  \\
$\Delta \Sigma$ & 0.409(34) & 0.402(48) & 0.318 \\
$2 \,L$ & 0.005(52) & 0.050(54) & 0.358 \\ \botrule
\end{tabular} \label{LHPC}}
\end{table}
  
Naturally, one must be careful about large uncertainties inherent in the
lattice QCD calculations at this stage. 
They suffer from various limitations, which come from the quenched
approximation, the finite-size effects of the lattice, large pion
mass effects and/or the ambiguities in the chiral extrapolation
procedures, etc.
Also noteworthy is the fact that, in the simulation by the LHPC and the
QCDSF-UKQCD groups, only the contributions of connected-insertion (CI)
were taken into account and those of the disconnected-insertion (DI)
were totally left out.

More recently, $\chi$QCD Collaboration carried out a challenging
study of nucleon spin contents by including the DI contributions
as well and found that they in fact have sizable effects \cite{Liu12,Deka13}.
Their results are shown in Table \ref{chiQCD}.
They confirmed the results by the LHPC and the
QCDSF-UKQCD collaborations that the CI contribution
to the net quark OAM is certainly very small. However, they
found that the DI contribution to the same quantity is very large.
As a consequence, their result shows that nearly half of the nucleon
spin comes from the quark OAM (``mechanical'' OAM). This number is
even larger than the prediction of the CQSM, although they are
consistent in a qualitative sense. In view of the previously-mentioned
various uncertainties of the lattice QCD calculation, it would be
premature to draw a decisive conclusion at the present stage.
Nevertheless, their analysis clearly reminds us of the fact that
some of the nucleon observables are very sensitive to the
introduction of the DI contributions, which are thought to simulate
the pion clouds effects dictated by the spontaneous
chiral symmetry breaking of QCD vacuum. This means that, in
oder to get realistic predictions for internal structures of the
nucleon in the framework of lattice QCD, more serious account of
the DI contributions is absolutely necessary.
Also highly desirable is
to carry out calculations with dynamical fermions.

\begin{table}[ph]
\tbl{Lattice QCD estimate of the contributions of connected insertions (CI)
and the disconnected insertions (DI) to the nucleon spin contents by
the $\chi$QCD Collaboration \cite{Liu12,Deka13}.}
{\begin{tabular}{@{}cccc@{}} \toprule
\hspace{18mm} & \ \ \ \ CI ($u+d$) \ \ \ \ & \ \ \ \ DI ($u+d+s$) \ \ \ \ 
& \ \ \ \ \ \ sum \ \ \ \ \ \ \\ \colrule
$2 \,J$ & 0.629(51) & 0.092(14) & 0.72(8) \\
$\Delta \Sigma$ & 0.62(9) & - \,0.36(3) & 0.25(12) \\
$2 \,L$ & 0.01(10) & 0.46(3) & 0.47(13) \\ \botrule
\end{tabular} \label{chiQCD}}
\end{table}
 
So far, our eyes are mainly turned on flavor singlet
combination (or the net contribution) of the quark OAMs.
Also very interesting is the isovector combination, i.e. the
difference of OAMs carried by $u$-quark and $d$-quark.
It should be emphasized that the lattice QCD predictions for
this flavor-nonsinglet quantity are expected to be quantitatively
more trustable than that for the flavor-singlet quantity $L^{u+d}$,
because there is no DI contribution to the former, which is harder
to estimate reliably. 
As already pointed out, the predictions of the LHPC and QCDSF-UKQCD
groups are that $L^u$ is negative and $L^d$ is positive, which leads to
a remarkable prediction that $L^u - L^d$ is sizably negative
\begin{equation}
 2 \,L^{u-d} \ \simeq \ - \,0.4.
\end{equation}
This must be a surprise. In fact, it sharply contradicts
the prediction of the familiar quark model like the MIT bag model.
To explain it, we first recall the nucleon spin sum rule obtained
within the familiar MIT bag model in both of the isoscalar and isovector
channels. They are given by
\begin{eqnarray}
 2 \,J^{u+d} \ &=& \ \Delta \Sigma^{u+d} \ + \ 2 \,L^{u+d} , \\
 2 \,J^{u-d} \ &=& \ \Delta \Sigma^{u-d} \ + \ 2 \,L^{u-d} ,
\end{eqnarray}
where $2 \,J^{u+d} = 1$, $2 \,J^{u-d} = 5/3$, while
\begin{eqnarray}
 \Delta \Sigma^{u+d} \ &=& \ \int_0^R \,\left\{\,[f(r)]^2 \ - \ 
 \frac{1}{3} \,[g(r)]^2 \,\right\} r^2 \,dr, \\
 \Delta \Sigma^{u-d} \ &=& \ \frac{5}{3} \,\Delta \Sigma^{u+d} ,
\end{eqnarray}
and
\begin{eqnarray}
 L^{u+d} \ &=& \ \frac{2}{3} \,\int_0^R \,[g(r)]^2 \,r^2 \,dr , \\
 L^{u-d} \ &=& \ \frac{5}{3} \,L^{u+d} .
\end{eqnarray}
Here, $R$ is the bag radius, while $f(r)$ and $g(r)$ are the radial
wave functions of the upper and lower components of the ground
state of the MIT bag model, given in the form : 
\begin{equation}
 \psi_{g.s.} (\bm{r}) \ = \ \left(\, \begin{array}{c}
 f(r) \,\chi_s \\
 i \,g(r) \,\bm{\sigma} \cdot \hat{\bm{r}} \,\chi_s \\
 \end{array} \,\right) ,
\end{equation}
with the normalization
\begin{equation}
 \int_0^R \,\left\{\,[f(r)]^2 \ + \ [g(r)]^2 \,\right\} \,r^2 dr \ = \ 1.
\end{equation}
Note that the limiting case of non-relativistic quark model is obtained by
setting the lower component to be zero, i.e. $g(r) \equiv 0$.
In this limit, we have
\begin{equation}
 \Delta \Sigma^{u+d} \ = \ 1, \ \ \ L^{u+d} \ = \ 0,
\end{equation}
and
\begin{equation}
 \Delta \Sigma^{u-d} \ = \ \frac{5}{3}, \ \ \ L^{u-d} \ = \ 0.
\end{equation}
This just reconfirms the fact that, in the MIT bag model, the quark OAMs come
from the lower $P$-wave component of the ground-state wave function.
For a typical bag parameter, which
gives $\Delta \Sigma^{u+d} \simeq 0.7$, one would find that
\begin{equation}
 2 \,L^{u+d} \ \simeq \ 0.30, \ \ \ 2 \,L^{u-d} \ \simeq \ 0.50,
\end{equation}
which especially means that $L^u - L^d$ is positive with sizable
magnitude. This is in sharp contradiction to the afore-mentioned
prediction of the lattice QCD, $2 \,L^{u-d} \simeq - \,0.4$.
We recall that the discrepancy between the EMC observation
$\Delta \Sigma \simeq (0.2-0.3)$ and the predictions of the standard
quark models (remember for example, the prediction of the naive quark
model $\Delta \Sigma^{u+d} = 1$, or the prediction of the MIT bag
model $\Delta \Sigma^{u+d} \simeq 0.7$) was called the ``nucleon
spin crisis''. Now, the discrepancy between the isovector combination
of the quark OAMs pointed out above seems more drastic, in the sense
that even their signs are different. One might then call it ``another''
nucleon spin crisis.

There can be two possible origins of this discrepancy.
The first is the possibility of significant numerical difference between
the two OAMs, i.e. the canonical and mechanical OAMs.
The OAM difference $L^u - L^d$
calculated above in the MIT bag model roughly corresponds to the
canonical OAM, although the MIT bag model is not a gauge theory.
On the other hand, the OAM $L^u - L^d$ extracted from the lattice QCD
analysis with help of the Ji sum rule is the mechanical OAM.
There is no reason to expect that their difference is small.

The second possible origin is the strong scale dependence of $L^u - L^d$,
which we shall discuss below. Remember first
that the renormalization scale of the lattice QCD calculation corresponds to
relatively high-energy scale as $Q^2 \simeq 4 \,\mbox{GeV}^2$.
On the other hand, the energy scale of low energy models like the
MIT bag model is believed to be much lower, say $Q^2 \simeq
(0.4 \sim 0.6 \,\mbox{GeV})^2$. This makes no big difference in the case
of flavor singlet quark spin, because it is nearly a scale-independent
quantity, except in the extremely low energy regions where the
framework of the perturbative renormalization group becomes
untrusted.\footnote{Here, we are supposing the
gauge-invariant $\overline{{\rm MS}}$ factorization scheme not the 
Adler-Bardeen scheme.}
However, the isovector combination of the quark OAMs turns out to be
strongly scale-dependent quantities \cite{WN06,WN08,Thomas08}.
Thomas then claims that this strong scale-dependence of
$L^u - L^d$ is likely to resolve the above-mentioned discrepancy at
least partially \cite{Thomas08,MT88}.
This explanation was criticized by Wakamatsu \cite{Waka10}, however.    
Thomas' analysis starts from an estimate of the $u$- and $d$-quark OAMs
based on the improved cloudy bag model which also takes account of
the exchange current contribution associated with the
one-gluon-exchange hyperfine interactions. Those model predictions
are regarded as initial scale values corresponding to
a very low energy scale, say, $0.4 \,\mbox{GeV}$.
Then, by solving the QCD evolution equation for the $u$- and $d$-quark
OAMs first derived by Ji, Tang, and Hoodbhoy \cite{JTH96}, he found that 
the OAMs of $u$- and $d$-quarks cross over around the scale of
$0.5 \,\mbox{GeV}$. This crossover of $L^u$ and $L^d$ is just what
is required from the consistency with the lattice QCD results given
at the scale of $Q^2 \simeq 4 \,\mbox{GeV}^2$.
(Actually, a careful observation reveals that the discrepancy between
Thomas' prediction and the lattice QCD predictions at the scale
$Q^2 = 4 \,\mbox{GeV}^2$ is fairly large.) 
As pointed out in Ref.\refcite{Waka10}, however, the starting energy
of evolution used in his analysis is fairly low, and, at such low energy scales,
$L^{u-d}$ shows tremendously strong scale dependence.
In fact, this behavior of $L^{u-d}$ is related to the diverging
behavior of the QCD running coupling constant $\alpha_S (Q^2)$ as
$Q^2 \rightarrow 0$.
If the magnitude of $\alpha_S$ becomes too large, one must suspect
the validity of the used QCD evolution equation, which is based on
the framework of {\it perturbative} renormalization group equation.
Also very difficult to know is the size of ambiguity arising from
the choice of the starting energy of evolution, because the renormalization
scale of any effective model can be given only by a crude guess.
Wakamatsu then advocated the following strategy.
Instead of carrying out upward evolution
by starting from the predictions of effective models corresponding
to low energy scales, one may start with the information known at
the high energy scales, say, at $Q^2 \simeq 4 \,\mbox{GeV}^2$ and
to carry out a downward evolution by leaving the question where to
stop this downward evolution.
The quantity $L^{u-d}$ at the scale of $Q^2 = 4 \,\mbox{GeV}^2$ can
be estimated by using the relation
\begin{equation}
 L^{u-d} \ = \ J^{u-d} \ - \ \frac{1}{2} \,\Delta \Sigma^{u-d} ,
\end{equation}
with
\begin{equation}
 J^{u-d} \ = \ \frac{1}{2} \,\left[\, \langle x \rangle^{u-d} 
 \ + \ B^{u-d}_{20} (0) \,\right] .
\end{equation}
Here, $\Delta \Sigma^{u-d}$ is identified with the beta-decay coupling
constant $g_A^{(I=1)}$ of the neutron, which is known with high precision.
The difference $\langle x \rangle^{u-d}$ between the $u$- and $d$-quark
momentum fractions at $Q^2 = 4 \,\mbox{GeV}^2$ is also a fairly precisely
known quantity from the global analysis of the inclusive DIS
data \cite{MST06,CTEQ00}.
Only one unknown is therefore the isovector anomalous gravito-magnetic
moment $B^{u-d}_{20} (0)$ of the nucleon. Fortunately, this is an
isovector quantity, which receives no DI contribution, so that one
can expect that the corresponding lattice QCD prediction by the LHPC or
the QCDSF-UKQCD collaborations is much more reliable than that
for $B^{u+d}_{20} (0)$.
The value of $L^{u-d}$ estimated in this way is used as a initial
condition given at $Q^2 = 4 \,\mbox{GeV}^2$, and the downward evolution
was carried out to obtain the value of $L^{u-d}$ at the low energy scales
corresponding to effective models of the nucleon.

\begin{figure}[h]
\begin{center}
\includegraphics[width=7.0cm]{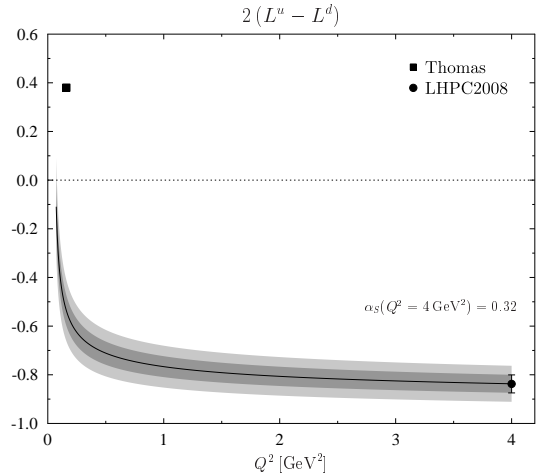}
\caption{\label{Lu-Ld}The scale dependence of $2 \,L^{u-d}$ obtained
by solving the leading-order evolution equation with the LHPC
lattice QCD prediction at $Q^2 = 4 \,\mbox{GeV}^2$ with errors
as initial condition of downward evolution. Also shown by the filled square
is the prediction of the improved cloudy bag model corresponding to
the scale $Q^2_0 = (0.4 \,\mbox{GeV})^2$.}
\end{center}
\end{figure}

Fig.\ref{Lu-Ld} shows an example of such calculations \cite{Waka10}.
One sees that the scale
dependence of $L^{u-d}$ is in fact quite strong as $Q^2$ becomes very low.  
At the unitarity violating limit ($Q^2 \simeq (0.225 \,\mbox{GeV})^2$),
$2 \,L^{u-d}$ becomes close to zero.\footnote{The unitarity violating limit here
means the scale where the gluon momentum fraction becomes {\it negative},
when one performs a downward evolution by starting from the empirical values
of quark and gluon momentum fractions given at $Q^2 = 4 \,\mbox{GeV}^2$.}
However, at the favorite matching scale with the cloudy bag model, i.e.
$Q^2 \simeq (0.4 \,\mbox{GeV})^2$, one still finds that $2 \,L^{u-d}$ is
sizably negative, i.e. $2 \,L^{u-d} \simeq - \,0.53$.
This negative value with large magnitude is in sharp contradiction with
the predictions of the refined cloudy bag model or any quark model with
SU(6)-like spin-flavor structure.
Very curiously, the prediction of the CQSM for
$2 \,L^{u-d}$ given in the papers \refcite{WT05,Waka10} is negative
with large magnitude, and
it is consistent with the above phenomenological estimate with partial
use of the lattice data. It would be interesting to point out the fact
that the theoretical prediction for $2 \,L^{u-d}$ given in these
papers is not the canonical OAM. It is obtained through the calculation
of the forward limit of $A^{u-d}_{20} (0)$ and $B^{u-d}_{20} (0)$
together with the Ji sum rule. It is therefore a quantity which is conceptually
closer to the mechanical OAM rather than the canonical OAM, although
the distinction between these two OAMs is not so clear for an effective
quark model like the CQSM, which is not a gauge theory.

Does the strong scale dependence of $L^{u-d}$ rescue the discrepancy
between the prediction of the lattice QCD and that of the low energy
models or not ? Or does it suggest a significant numerical difference
between the canonical and mechanical OAMs in the nucleon ?
The answer to this question is not yet absolutely clear.
However, it seems clear that the above mentioned puzzle provides us
with valuable nontrivial information on the role of quark OAMs in
the nucleon spin decomposition problem.

\section{Summary and concluding remarks}

Now we are in a position to answer the proposed question in the
present paper. ``Is gauge-invariant complete decomposition of the
nucleon spin possible ?''\footnote{The gauge-invariance here should be
taken as a traditional one, not as the weak gauge-invariance {\it a la}
Lorc\'{e}.}
The truth appears that this question is a little bit too general to give a unique
answer. If the question concerns the most general nucleon spin decompositions
including the transverse spin sum rules, the answer is likely to be ``No''.
On the other hand, if the question concerns the most fundamental
longitudinal nucleon spin decomposition, the answer would
most probably be ``Yes''. The reason is the following. 
The two ``seemingly'' covariant gauge-invariant decompositions (I)
and (II) of the nucleon spin proposed by Wakamatsu is of
general nature in the sense that it still has a large degrees freedoms
such that it can be reduced to any known gauge-invariant decompositions
after an appropriate choice of the Lorentz frame of
reference. In particular, there is no doubt about that the decomposition (II)
contains the two popular gauge-invariant decompositions of Chen et al. and
of Bashinsky-Jaffe, depending on an appropriate choice of
Lorentz frame and a suitable condition which is necessary to uniquely
specify the decomposition of the gauge field into the physical and pure-gauge
components. Since each term of those two decompositions is separately
gauge-invariant, both are clearly gauge-invariant decompositions.
Remember the fact that, in the QED case, the Chen decomposition
is nothing but the familiar transverse-longitudinal decomposition of
the photon field and in particular that the transverse component
is a gauge-invariant quantity with unambiguous physical meaning.
Still, things to be worried about here is that the transverse-longitudinal
decomposition or the concept of transversality of the gauge
field is generally Lorentz-frame dependent concept.
This is the reason of general statements found in many standard
textbooks of electrodynamics, which tells that the total photon angular
momentum cannot be gauge-invariantly decomposed into the orbital and
intrinsic spin parts.
This statement would certainly be true in the most general
context. However, as shown in sect.5, this is not necessarily the case
for the longitudinal component of the total photon angular momentum.
The point is that the helicity for a massless particle is a Lorentz-invariant
quantity.\footnote{In fact, there is an explicit proof that
the photon helicity is an invariant, even though, in general, Lorentz
boosts transform the transverse, longitudinal, and the time-like
components of the vector potential into each other \cite{HS78}.} 
The component of the total photon angular momentum along
the direction of the photon momentum can be decomposed into the orbital
and intrinsic spin parts in a gauge- and frame-independent way and both
are definite observables. 

Coming back to our nucleon spin decomposition problem, the helicity
sum rule of the nucleon is basically a Lorentz-frame independent sum rule.
In particular, it is invariant under a wide class of Lorentz boost
in the direction of the nucleon momentum\footnote{Naturally, since
the nucleon is a massive particle, its helicity can
change under an extremely fast Lorentz boost in the nucleon momentum
direction.} .     
Because the longitudinal spin sum rule of the nucleon, or the helicity sum
rule, is invariant under such Lorentz-boosts, one can work in any
Lorentz-frame. This especially means that there is nothing wrong in
working within a noncovariant framework like in the Chen decomposition.
A logical conclusion drawn from this consideration is that the Chen
decomposition and the Bashinsky-Jaffe decomposition (or the Hatta
decomposition), which can be reduced from more general decomposition (II)
{\it a la} Wakamatsu would give the same answer for the longitudinal
decomposition of the nucleon spin. Note, however, that this is not true for
more general nucleon spin decompositions like the transverse decomposition
of the nucleon spin. At any rate, an important conclusion drawn from the
consideration above is that the longitudinal gluon spin $\Delta G$ is
most likely to be a gauge- and frame-independent observable.
In other words, $\Delta G$ is a gauge-invariant quantity in a traditional
or strong sense at variance with the statement in the review by
Leader and Lorc\'{e} \cite{Leader_Lorce13}. 

Another important subject addressed in the present review is the question
of observability of the two kinds of OAMs of quarks and gluons,
i.e. the mechanical OAM and the generalized (gauge-invariant) canonical OAM.
Now it is a wide-spread belief in the QCD spin physics community that
the canonical OAM (not the mechanical
OAM) is the quantity with natural physical interpretation as OAMs of
free partonic motion of constituents, i.e. quarks and gluons.
There are two reasons for this belief. First, the generalized canonical
OAMs are believed to obey the standard angular momentum commutation
relation, so that they are supposed to work properly as generators of
spatial rotation. 
Second, it is widely believed that the dynamical quark OAM is given as a sum
of the canonical quark OAM and the potential angular momentum, which
appears to support the interpretation that the dynamical quark OAM
contains the genuine twist-3 quark-gluon interaction term.
As explained in the present paper, both these beliefs are not necessarily
justified. Concerning the first one, we have shown that, for a massless
photon, neither of the canonical OAM nor the intrinsic spin satisfies
the SU(2) algebra. It was also shown that this observation is inseparably
connected with the fact that there is no rest frame for a massless
particle.  
Concerning the second question, we have given a plausible argument to
show that what contains the potential angular momentum is rather the
canonical quark OAM than the dynamical quark OAM, in contradiction to
naive expectation. These observations are by no means academic ones.
In fact, they have important consequences on the possible observability
of the two OAMs. Now we know that the canonical OAM is related
to a certain moment of the Wigner distribution, which is expected to
describe the partonic orbital motion of quarks in the plane perpendicular
to the direction of nucleon momentum and spin. However, the recent
paper by Courtoy et al. revealed a principle difficulty of observing the
relevant distribution appearing in this sum rule \cite{CGHLR13}.
On the other hand, the dynamical OAMs are already known to have clear
relations to DIS observables. One is the indirect relations through
the GPDs and the Ji sum rule. The other is the relation in which
the dynamical quark OAM is given as a 2nd moment of the GPD $G_2$.
Although the actual experimental determination of $G_2$ would not be an
easy task, an interesting fact is that the genuine twist-3 part of
$G_2$ does not contribute to this sum rule and that the Wandzura-Wilczek
part of $G_2$ is completely determined by the twist-2 GPDs $H^q (x,0,0)$,
$E^q (x,0,0)$, and $\tilde{H}^q (x,0,0)$.
This appears to support our viewpoint that what contains the genuine
twist-3 quark-gluon interaction term is not the dynamical quark OAM
but the canonical quark OAM.

To sum up, what descriminates the two gauge-invariant decompositions of
the nucleon spin are the orbital angular momentum parts of quarks and gluons.
They are specified by the two different OAMs, i.e. the dynamical OAM and
the canonical OAM. For a weakly-coupled gauge system like the hydrogen
atom, there is no practical difference between these two OAMs
and there is nothing wrong in believing that the canonical OAM is
a natural building block of quantum theory.
This is because the weak interactions between the transverse photons and
the charged particles can be introduced and handled at later stage
as a perturbation.
For a strongly-coupled gauge systems like the nucleon, however, the
distinction between the canonical OAM and the dynamical OAM becomes crucial.
Here, we cannot neglect the Fock-components of the transverse gluon
in the nucleon wave function.
Otherwise, we would have no gluon distributions.
This means that, when one talks about
the OAMs of quarks and gluons in the nucleon, one must at the least be
clearly conscious of which OAMs one is thinking of.

\section*{Acknowledgments}

The author greatly acnowledge numerous enlightning discussion with
E. Leader and C. Lorc\'{e}. He also grealy appreciates useful discussions
with R.L.~Jaffe, M. Burkart, S.C. Tiwari, Y. Hatta, S. Yoshida, H.-Y. Cheng,
F. Wang, X.S. Chen, K.-F. Liu, S. Liuti, and T. Kubota.

\end{document}